\documentclass{article}[12pt]
\usepackage{epsf}
\usepackage[latin1]{inputenc}
\usepackage{amsmath,amsfonts,amssymb}
\usepackage{graphicx,xspace}
\usepackage{array}
\def\beq{\begin{equation}}
\def\eeq{\end{equation}}
\def\bea{\begin{eqnarray}}
\def\eea{\end{eqnarray}}

\def\neq{\not=}

\def\inv{^{-1}}

\def\tilde{\widetilde}
\def\bar{\overline}

\def\cech{${\rm C}^{\kern-6pt \vbox{\hbox{$\scriptscriptstyle\vee$}\kern2.5pt}}${\rm ech}}
\def\Cech{${\sl C}^{\kern-6pt \vbox{\hbox{$\scriptscriptstyle\vee$}\kern2.5pt}}${\sl ech}}

\def\CE{{\cal E}}
\def\CF{{\cal F}}

\def\CI{{\cal I}}

\def\CP{{\cal P}}

          \def\CP{{\cal P}}

\def\inv{^{-1}}

\def\inv{^{\raise.15ex\hbox{${\scriptscriptstyle -}$}\kern-.05em 1}}

\def\({\left(}
\def\){\right)}
\def\<{\left\langle\,}
\def\>{\, \right\rangle}
\def\[{\left[}
\def\]{\right]}

\newcommand{\newsec}[1]{\setcounter{equation}{0} \section{#1}}

\def\E{{\cal E}}
\def\bl{\[}
\def\br{\]}

\date{}

\begin{document}

\begin{titlepage}
\hfill {IPhT t11/197}
\bigskip
\bigskip
\begin{center}
{\large\bf
Conformal blocks in Virasoro and W theories: duality and the Calogero-Sutherland model
}\\[.3in] 

{\bf Benoit Estienne$^{1}$,   Vincent Pasquier$^{2}$, Raoul\ Santachiara$^{3}$ and Didina Serban$^{2}$}\\
	$^1$ {\it Institute for Theoretical Physics, Universiteit van Amsterdam \\
Valckenierstraat 65, 1018 XE Amsterdam, The Netherlands\\
            e-mail: {\tt b.d.a.estienne AT uva.nl}}\\
         
          $^2$ {\it Institut de Physique Th\'eorique, DSM, CEA, URA2306 CNRS, Saclay, F-91191 Gif-sur-Yvette, France \\ e-mail: {\tt vincent.pasquier AT cea.fr; didina.serban AT cea.fr}}\\

	$^3$   {\it Laboratoire J.~V.~Poncelet, UMI 2615, Moscow and \\ LPTMS,CNRS,UMR 8626,
             Universit\'e Paris-Sud,\\ 
             B\^atiment 100,
             91405 Orsay, France. \\
    e-mail: {\tt raoul.santachiara AT lptms.u-psud.fr}. }\\
\end{center}
\vskip .04in
\centerline{(Dated: \today)}
\vskip .2in
\bigskip
\centerline{\bf ABSTRACT}
\begin{quotation}
We study the properties of the conformal blocks of the conformal field theories with Virasoro or W-extended symmetry. When the conformal blocks contain only second-order degenerate fields, the conformal blocks obey second order differential equations and they can be interpreted as ground-state wave functions of a trigonometric Calogero-Sutherland Hamiltonian with non-trivial braiding properties. 
A generalized duality property relates the two types of second order degenerate fields. By studying this duality we found that the excited states of the Calogero-Sutherland Hamiltonian are characterized by two partitions, or in the case of $\textrm{WA}_{k-1}$ theories by $k$ partitions. By extending the conformal field theories under consideration by a $u(1)$ field, we
find that we can put in correspondence the states in the Hilbert state of the extended CFT with the excited non-polynomial eigenstates of the Calogero-Sutherland Hamiltonian. 
When the action of the Calogero-Sutherland integrals of motion is translated on the Hilbert space, they become identical to the integrals of motion recently discovered by Alba, Fateev, Litvinov and Tarnopolsky in Liouville theory in the context of the AGT conjecture. Upon bosonisation, these integrals of motion can be expressed as a sum of two, or in general $k$,
bosonic Calogero-Sutherland Hamiltonian coupled by an interaction term with a triangular structure. For special values of the coupling constant, the conformal blocks can be expressed in terms of Jack polynomials with pairing properties, and they give electron wave functions for special Fractional Quantum Hall states.
\end{quotation}
\end{titlepage}

\vskip 0.5cm
\noindent

\newsec{Introduction}

Since their introduction forty years ago \cite{Ca,Su}, the Calogero-Sutherland  models, which describe one-dimensional  particles interacting via pairwise inverse square potential,   have gained considerable interest in theoretical  and mathematical  physics.   The classical  \cite{Ols_Per} and quantum \cite{UHW,Poly}  Calogero-Sutherland systems have been proven to be completely integrable and the algebraic structures responsible for the solvability of these models have appeared in various area of theoretical physics, such as  random matrix  theories or  two dimensional Yang-Mills theories. Moreover  the Calogero-Sutherland systems have been shown to  belong  to a wide class of fully solvable systems  associated to Lie algebras (see \cite{Dunkl_rev, Dunkl_Opd}) which are relevant for the study of  orthogonal polynomials associated to Lie root lattices.  

Here  we are interested into  the connection between CS models and  2d conformal field theories (CFTs) which are based on Virasoro or, more generaly, on  $\textrm{WA}_{k-1}$ algebras. The latter  are generated  by $k-1$ conserved  currents of spin $s=2,..,k$ and their representations are associated to the root lattice $A_{k-1}$.  The $k=2$  case corresponds   to the Virasoro algebra.  In the following,  we will understand by Calogero-Sutherland  model the quantum trigonometric version which describes particles on a ring.  For this model, the exact evaluation of the ground state correlations, both static and dynamic, has been possible \cite{Ha, SLP} by using the theory of Jack polynomials. 

As it was observed in  \cite{Forr, AMOS} by considering  the properties of  a class of multidimensional integrals, the Selberg-Aomoto integrals \cite{Kaneko}, the CS model is  intrinsically related to the Virasoro algebra $Vir(g)$ with central charge $c=1-6(g-1)^2/g$. Here, $g$ parametrizes the CS coupling, 
and the central charge $c$ is invariant under the change $g \to 1/g$ which corresponds to the duality transformation of the CS model  \cite{Stanley,Gaudin, Macdonald}. 
One aspect of this  connection is that  the Jack polynomials  characterize the  Virasoro  singular vectors \cite{MY,SSDFR}. Analogous results hold for generic $\textrm{WA}_{k-1}$ algebras \cite{AMOS_w}.

Another aspect of the relationship between $Vir(g)$  CFTs and  CS$(g)$ model has been made  explicit  in the context  of   Schramm-Loewner evolutions (SLE) \cite{Cardy}.  The probability measure associated to the evolutions of $N$ SLE traces is given by  the conformal blocks involving  $N$ second order degenerate fields, i.e. Virasoro primaries which possess a second level null vector in their Verma module. The conformal blocks  involving these degenerate fields satisfiy a second order differential equation  which can  be related to the  CS Hamiltonian \cite{Cardy,Cardy_Doyon}. 
Consequently, these conformal blocks form a new  family of  CS eigenfunctions.   As  we will discuss more in detail later, these eigenfunctions are in general not polynomial (they do not correspond to Jack polynomials) and are characterized by a non-abelian monodromy.    

These findings were extended  to  the conformal blocks of $N$  primaries fields of a general  $\textrm{WA}_{k-1}$ theory \cite{ES}.    The main motivation in \cite{ES} was the study of  a class of trial many-body wavefunctions for fractional quantum Hall effect (FQHE) at bosonic filling fraction $\nu=k/r$ \cite{ES,EBS}. These states can be constructed  from  the   conformal blocks of a series of  minimal models of $\textrm{WA}_{k-1}$ algebras, the $\textrm{WA}_{k-1}(k+1,k+r)$ theories. This is the case for instance for  the  Moore-Read states \cite{MooreRead,Nayak_Wilczek} with $k=2$, $r=2$  and for the Read-Rezayi states \cite{ReadRezayi} with $k>2, r=2$ which play a paradigmatic role in the physics of non-abelian FQHE states.  For a given $\textrm{WA}_{k-1}$ algebra, {\it i.e.} for a given central charge, there are two primary fields, which we indicate as $\Psi$ and $\sigma$ fields\footnote{We borrow the notations  used in FQHE for, respectively,  the electron and quasihole operator.},  whose representation modules present the same degeneracy structure and whose correlation functions satisfy second order differential equations.  In the CFT approach to FQHE,  a quasihole wavefunction is given by a conformal block involving $\Psi$ and $\sigma$ fields. In \cite{EBS} it was found that  the differential operator which annihilates  the quasihole wavefunction  decomposes as the sum of two independent CS models  acting separately on the  coordinates of the $\Psi$ and of the  $\sigma$ fields.  These two CS Hamiltonians have dual coupling strength  \cite{EBS}. In this sense, the conformal block of $\Psi$ and  $\sigma$ fields generalizes the duality kernel \cite{Stanley, Macdonald, Gaudin} in the non-abelian setting.  By studying the edge excitations  for non-abelian quantum Hall states,  it has been shown in \cite{KJS,KJS_2} that this duality manifests also in the CFT characters  characterizing the quasihole and particle sectors.

The purpose of this paper is to use the separation of variables mentioned above to study in more detail the new structures of the CS conformal block eigenfunctions. In this respect  we are  lead to consider the CFT which is based on  the tensor product  of  the Heisenberg algebra and the $\textrm{WA}_{k-1}$ algebra, $u(1)\otimes \textrm{WA}_{k-1}$. We will show that a conformal block of primary operators giving  a  CS eigenfunction plays the role of  a reference ground state upon which we  can define  a family  of  CS excited eigenstates. These eigenstates   are  indexed by  $k$ Young tableaux and   are obtained by inserting  into the conformal block  a particular  $u(1)\otimes \textrm{WA}_{k-1}$ descendant field.  We will prove that the descendant states associated to the CS eigenfunctions  form an  orthogonal basis which, in the case of the $u(1)\otimes Vir(g)$ algebra,   corresponds to the basis introduced  in \cite{AFLT} to investigate the AGT conjecture \cite{AGT}, i.e.  the  expansion  of the  conformal  blocks of Liouville theory (or more generally of CFTs based on Virasoro algebra \cite{FL,ST}) in terms of Nekrasov instanton functions \cite{Nekrasov} of $SU(2)$ gauge theories\footnote{Our results can be related to those in \cite{AFLT} by changing $g\to -g$.}.  In some sense, our finding is not surprising.  The  basis of  descendant states have been shown \cite{AFLT} to diagonalize  a series of commuting integrals of motion which, in the classical limit, correspond to the  Benjamin-Ono  conserved quantities. The Benjamin-Ono integrable hierarchy is  in turn related to the  CS system \cite{ABW}. The advantage of our approach is that the connection between the CS Hamiltonian and the set of commuting integral of motion obtained in \cite{AFLT} is direct and explicit. For general $k$, the basis of descendant states associated to CS eigenfunctions is expected \cite{AFLT} to play an analogous role  in  the generalization of the AGT conjecture concerning the $SU(k)$ gauge theories and the $\textrm{WA}_{k-1}$ theories \cite{W,MM}.

Integrability of the CS Hamiltonian implies that there exists other, higher order conserved quantities which should be simultaneously diagonalized by the CS eigenfunctions. We have proven that the conformal blocks discussed above also obey a third-order differential equation which is related to the third order CS Hamiltonian. In order to prove this property, we have used the null vector condition for a particular descendant of the second order null vector. 
In our approach, it is still an open question how to systematically obtain the whole tower of integral of motion. We have verified that our third order 
integral of motion coincides with the expression $I_4$ conjectured in Appendix C of \cite{AFLT}.

Our strategy is as follow: we first conjecture the eigenenergy formula for the CS non-polynomial eigenstates by using the separation of the variables between $\Psi$ and $\sigma$ fields and the singularity structure of the conformal blocks. Second, we translate the action of the CS Hamiltonian in the differential form into an operatorial form similar to that of \cite{AFLT}, involving the Heisenberg and Virasoro (or more generally $\textrm{WA}_{k-1}$) generators. Then, using the bosonisation of Virasoro ($\textrm{WA}_{k-1}$) algebra, and performing a change of basis in the space of bosonic fields, like in \cite{AFLT} and in \cite{BB} we write the integral of motion associated to the CS Hamiltonian in terms of $k$ bosonic fields. As shown by Belavin and Belavin  \cite{BB} in the case $k=2$, for $g=1$ the CS Hamiltonian splits into $k$ copies of one-component bosonized CS Hamiltonians \cite {Jevicki}, with a trivial coupling term involving the zero modes. This splitting explains in particular why we  can characterize the generic CS eigenstates using $k$ partitions. Outside the point $g=1$, the CS Hamiltonian is a sum of $k$ copies of one-component bosonized CS Hamiltonians with $g\neq1$, plus a coupling term with a triangular structure in the creation/annihilation operators. The triangular structure of the coupling term insures that the spectrum is still given by the sum of $k$ one-component CS eigenenergies, each characterized by a partition. This proves our initial conjecture on the eigenenergies.
Finally, the duality $ g \to 1/g$ has a very simple realization in terms of the bosonized CS Hamiltonians and it gives rise to two dual bases in the Hilbert space.

The paper is organized as follows: sections 2 and 3 are reviewing basic facts about the CS model and CFT's respectively and are fixing the notations. 
Since the generic formulas for the $\textrm{WA}_{k-1}$ algebras are rather complicated, we have preferred to treat first the Virasoro case, $k=2$ in section 4 and then repeat the computation for generic $k$ in section 5. In Appendix A we specialize to the Ising case, which is a special case where a class of eigenfunctions become polynomial. Appendix B contains details on the derivation of the operatorial form of the CS integrals of motion and Appendix C is devoted to the Coulomb gas representation of the non-polynomial CS eigenfunctions.

\newsec{Calogero-Sutherland model}
\label{Jacks}

In this section we review the standard relation between the Calogero-Sutherland model and Jack polynomials, and introduce some notations. 

\subsection{Integrability and Hamitlonians}

The trigonometric version of Calogero-Sutherland model is a one-dimensional quantum  model  defined by the  Hamiltonian:
\begin{equation}
H_{CS}^{g}\equiv-\frac{1}{2} \sum_{i=1}^{N} \frac{\partial^2}{\partial {x_i} ^2} + g(g-1) \sum_{1 \leq i<j\leq N}\frac{(\pi/L)^2}{\sin^2 \left[\pi(x_i-x_j)/L\right]}.
\label{CS_tri}
\end{equation}
The Hamiltonian (\ref{CS_tri}) describes a system of $N$  particles at positions   $x_i\in [0,L]$, $i=1,\dots ,N$  on a circle of perimeter  $L$ which  interact with  a long-range potential with coupling $g(g-1)$. It proves convenient to introduce the variables
\begin{equation}
 z_j=e^{2 i \pi x_j/L},
 \label{zvar}
\end{equation}  
in which the Hamiltonian (\ref{CS_tri}) takes the form (up to a multiplicative factor)
\begin{align}
H^{g}= \sum_{i=1}^{N} (z_i \partial_{i})^2 - g(g-1) \sum_{i\neq j}\frac{z_i z_j}{z_{ij}^2} \label{CS_g_00},
\end{align}
where $\partial_i = \partial / \partial z_i$ and $z_{ij} = z_i -z_j$. The total momentum operator reads as
\begin{equation}
\mathcal{P}=\sum_{i=1}^N z_i \partial_i.
\label{momoper}
\end{equation} 
The Calogero-Sutherland model is completely integrable \cite{BGHP}. The total momentum $\mathcal{P}$ and the Hamiltonian $H^g$ belong to a set of  $N$ functionally independent commuting operators $H_n^g$, whose first members are
 \begin{align}
H_1^g &= \mathcal{P} = \sum_{i=1}^{N} z_i \partial_{i} \\
H_2^g &= H^{g}= \sum_{i=1}^{N} (z_i \partial_{i})^2 - g(g-1) \sum_{i\neq j}\frac{z_i z_j}{z_{ij}^2}
\label{CS_g_0} \\
H_3^g &= \sum_{i=1}^N (z_i \partial_{i})^3 + \frac{3}{2}g(1-g)\sum_{i\neq j} \frac{z_iz_j}{z_{ij}^2}( z_i \partial_i - z_j \partial_j). 
\end{align}
These integral of motions, and the underlying integrability structure, can be derived using the so called Dunkl operator \cite{Dunkl_rev}.
In order to compare the Hamiltonian \eqref{CS_g_00} with expressions coming from CFT, it is particularly useful to conjugate with a generic Jastrow factor : 
\begin{align}
H^{g} \to H^{g,\gamma}= \Delta^{-\gamma} H^{g}  \Delta^{\gamma} & & \Delta^{\gamma}(z)=  \prod_{i<j} (z_i-z_j)^{\gamma}.
\label{sim_transf}
\end{align}
Naturally, if a function  $\Psi(z)$ is eigenvector of $H^{g}$, the function $\Delta^{-\gamma}(z)\Psi(z)$ is an eigenvector of $H^{g,\gamma}$ with the same eigenvalue. 
Under such a transformation the momentum $\mathcal{P}$ is simply shifted, and the Hamiltonian becomes
\begin{equation}
H^{g,\gamma}\equiv \sum_{i=1}^{N} (z_i \partial_{i})^2 +2(\gamma-g)(g+\gamma-1)\sum_{i< j}\frac{z_i z_j}{z_{ij}^2}+\gamma \sum_{i<j}\frac{z_i+z_j}{z_{ij}}(z_i\partial_i-z_j\partial_j)
\label{cs_gamma}
\end{equation}
up to a commuting term $a(g,\gamma) + b(g,\gamma) \mathcal{P}$, which is irrelevant for our purposes.  When studying the connection between Calogero-Sutherland wavefunctions and CFT correlator  we can safely consider the  $z$ variables as general  complex variables on the plane. Operators of the form \eqref{cs_gamma} will be used later. 

\subsection{Calogero-Sutherland eigenfunctions :  Jack polynomials and beyond}

Whenever two coordinates $z_i$ and $z_j$ approaches each other,  an eigenfunction $\Psi(z)$ of the Calogero-Sutherland Hamiltonian \eqref{CS_g_00} has a regular singularity 
\begin{align}
\Psi(z) \sim (z_i-z_j)^{\gamma} \label{bcs}
\end{align}
where the singular exponent $\gamma$ can only assume two possible values, namely $\gamma=g$ and  $\gamma=1-g$. Imposing the behavior of the wavefunctions when two particles collide is equivalent to choosing a particular  boundary conditions and thus fixing the Hilbert space in which the operator  (\ref{CS_tri}) acts. For instance in  the case  of  a repulsive interaction ($g>1$) between $N$ identical particles,  it is natural to select wavefunctions  behaving as $(z_i-z_j)^g$  for every couple of particles (note that for $g\geq 3/2$ this condition is  a necessary one to ensure  normalizability of the wavefunctions). 

The usual wave-functions of this model are obtained by imposing  the same boundary conditions for every couple of particles, and are of the form
 \begin{align}
\Psi^+(z) = \Delta^g(z) F^+(z) & & \Psi^-(z)  = \Delta^{1-g}(z) F^- (z)
\label{eig_bc}
\end{align}
where $F^{\pm}(z)$ are analytic functions when $z_i \to z_j$ and $\Delta^{\gamma}(z)$ is the Jastrow  factor \eqref{sim_transf}. It follows from  \eqref{cs_gamma} that these functions are eigenvectors of the so-called Laplace-Beltrami operator
 \begin{equation}
 \mathcal{H}^{\alpha}=\sum_{i=1}^{N} (z_i \partial_{i})^2 +\frac{1}{\alpha} \sum_{i<j}^{N}\frac{z_i+z_j}{z_{ij}}(z_i\partial_i-z_j\partial_j),
\label{CS_g_g}
\end{equation}
for $\alpha = 1/g$ and  $\alpha=1/(1-g)$, respectively\footnote{In the following we are using mixed notations to denote the CS Hamiltonians and their eigenfunctions, 
using the index $\alpha$ from the mathematical literature
to denote the Laplace-Beltrami form of the CS Hamiltonian (\ref{CS_g_g}) and its eigenfunctions, including the Jack polynomials, while we use the index $g$ for the untransformed Hamiltonians 
(\ref{CS_g_0}) and their eigenfunctions.}. The simplest  eigenfunctions of this type are the symmetric polynomials known as Jack polynomials $J^{\alpha}_{\lambda}$. They are labelled by partitions, \emph{i.e.} a decreasing sequence of positive integers $\lambda=[\lambda_1,\lambda_2\dots \lambda_N]$, and have eigenvalue 
\begin{equation}
\E_{\lambda}^{\alpha}=\sum_{i}^N \lambda_i\left[\lambda_i+\frac{1}{\alpha}(N+1-2i)\right].
\label{CS_eig_for}
\end{equation}
For more details on Jack polynomials we refer the reader to \cite{Macdonald}. This method allows to construct two branches of eigenfunctions for the Calogero-Sutherland Hamitlonian \eqref{CS_g_00}
\begin{align}
\Psi^+_{\lambda}(z) = \Delta^g(z) J^{1/g}_{\lambda}(z) &  & \Psi^-_{\lambda}(z)  = \Delta^{1-g}(z) J^{1/(1-g)}_{\lambda}(z)
\label{eig_cs_j}
\end{align}
for which all pair of particles have the same boundary conditions as they approach each other. Such wavefunctions can be interpreted as describing particles with abelian fractional statistics in the sense of Haldane \cite{Haldane_Statistics}.

However, as it has been well discussed in  \cite{Cardy_Doyon},  one can allow for more general boundary conditions, thus enlarging the Hilbert space under consideration.  New eigenstates  of  (\ref{CS_tri})  are  shown  to be given by certain conformal block of CFTs.  As we will discuss later, these new solutions are characterized by non-Abelian monodromies, and therefore can be thought of as describing non-Abelian anyons. This is precisely the type of  wavefunctions appearing in the context of non-Abelian states  \cite{MooreRead,ReadRezayi} in the fractional quantum Hall effect, as was obtained in \cite{EBS}.  These solutions form non-trivial representations of the braid group, and  therefore induce huge degeneracies in the  spectrum of the (\ref{CS_tri}) operator in these sectors.  In this paper we studied in full details these new solutions and we showed some clear and deep connections between the Calogero-Sutherland model and the integrable structure of the CFT. 

Moreover it is well known \cite{Stanley, Macdonald, Gaudin} that there is a duality relating the Calogero-Sutherland models (\ref{CS_tri}) with parameter $g$ and $1/g$. In particular this duality  relates the corresponding Jack polynomials through the decomposition of $\prod_{i,j}(1+z_i w_j)$ separating the variables $z_i$ and $w_j$:
\begin{equation}
\prod_{i,j}(1+z_i w_j)=\sum_{\lambda} J_{\lambda}^{1/g}(z) J_{\lambda'}^{g}(w), \label{separation_variables}
\end{equation}  
 where $\lambda^{'}$ is the transpose of $\lambda$.  We studied how this  duality manifests itself in this larger class of non-Abelian Calogero-Sutherland eigenfunctions described by certain CFT correlators, and we developed a method to tackle the problem of separating variables for these functions. 

 Before studying these non-Abelian eigenfunctions, we need to introduce some basic concepts of CFT which are behind this connection.

\newsec{CFT: basic notions }
\label{CFT_section}
We briefly review here the basic notions of CFTs. For a more in-depth introduction to CFT we refer the curious reader to \cite{diFrancesco}.

The  CFT is a two dimensional  quantum filed theory which enjoys conformal symmetry.  The CFT approach aims to compute  the correlator $\langle\Phi(z_1,\bar{z}_1),\dots \Phi(z_N,\bar{z}_N)\rangle$ 
of local fields  $\Phi(z,\bar{z})$ 
by exploiting the infinite number of constraints which the conformal symmetry in two dimension imposes.

\subsection{Virasoro algebra and primary fields}
The conformal symmetry  implies the existence of an holomorphic $T(z)$ and anti-holomorphic $\bar{T}(\bar{z})$ stress energy tensor. In two dimensions  the conformal group is the tensor product of holomorphic and antiholomorphic Virasoro algebras which are formed respectively by the Virasoro operators  $L_{n}$ and $\bar{L}_{n}$. For our purposes we consider  only the holomorphic part of the theory, \emph{i.e.} the holomorphic part of functions and fields.

The Virasoro operators $L_n$ are defined from the Laurent series of the stress-energy tensor $T(z)$
\begin{equation}
T(z) \Phi(w)\equiv \sum_{n}\frac{1}{(z-w)^{n+2}}L_{n} \Phi(w)
\end{equation}
and obey the commutation relations
\begin{equation}
[L_n,L_m] = (n-m)L_{n+m} + \frac{c}{12}n(n^2-1)\delta_{n+m,0}. 
\label{viras}
\end{equation}
The above relations define the Virasoro algebra with central charge  $c$.

 A Virasoro primary field $\Phi_{\Delta}(z)$  satisfies the following relations 
 \begin{equation}
   L_{0}\Phi_{\Delta}=\Delta \Phi_{\Delta} \quad L_{n}\Phi_{\Delta}=0 \;\;\mbox{for} \;\;n>0 
   \end{equation}
The $\Delta$ appearing in the above expression is the conformal dimension of the field primary $\Phi$. To each primary field correspond an infinite  family of fields, called descendants,  which are obtained by acting with the Virasoro operators on $\Phi_{\Delta}$, 
\begin{equation}
 \Phi_{\Delta}^{(n_1,n_2,\dots ,n_k)}= L_{-n_k}\dots L_{-n_1} \Phi_{\Delta}.
 \end{equation} 
 The descendant fields $\Phi_{\Delta}^{(n_1,n_2,\dots ,n_k)}$  are eigenvectors of $L_0$ with eigenvalue $\Delta+L$, where $L=\sum_i n_i$ is 
 called  level and classify the descendant fields.   For general values of $c$ and $\Delta$, all the independent fields are obtained by setting  
 $n_1\geq n_2\geq n_2\dots \geq n_k$. The number of possible descendants at a level $L$ is then equal to the possible partitions of $L$.
 
 \subsection{Degenerate fields and differential equations}
 For special value of the conformal dimension $\Delta$, $\Delta=\Delta(c)$, one can establish the existence of a descendant field $\chi(\Delta,L)$ at a certain  level $L$ such that $L_n \chi(\Delta,L)=0$ for $n>0$.
 The primary field $\Phi_{\Delta}$ is then said to be degenerate at level $L$ with $\chi(\Delta,L)$ being coined a null-vector. 

It is convenient in this respect  to parametrize the theory according to:
\begin{equation}
\label{min_mod_param}
c=1-6\frac{(g-1)^2}{g}
\end{equation}
and we denote the corresponding Virasoro algebra by $\textrm{Vir}(g)$. Trivially changing $g \to 1/g$  leaves the algebra invariant. As it will be clear later, the fact that we use  the same notation $g$ for the parameter fixing the central charge in the above expression and the coupling of the Calogero-Sutherland model in (\ref{CS_g_0}) is not casual.

Degenerate primary fields $\Phi_{(r|s)}$ are labelled by two integers $r$ and $s$. Their conformal dimension is
 \begin{equation}
 \Delta_{(r|s)}= \frac{1}{4} \left( \frac{r^2-1}{g} + (s^2-1)g + 2(1-rs) \right) \, ,
 \end{equation}
and they have a null-vector at level $L=r s$. Such a null vector is equivalent to a linear relation between usually independent  descendants. The identity operator, for instance, can be identified with the field $\Phi_{(1|1)}$ 
which presents a null-vector at level $L=1$
\begin{equation}
L_{-1}\Phi_{(1|1)}(z)=\partial_{z}\Phi_{(1|1)}(z)=0\;.
\end{equation}
Of particular interest are the operators $\Phi_{(1|2)}$ and $\Phi_{(2|1)}$, with conformal dimension
\begin{equation}
\Delta_{(1|2)}=\frac{3g -2}{4}\;, \qquad \Delta_{(2|1)}=\frac{3-2g}{4 g}\, .
\end{equation}
They are degenerate at level $L=2$:
\begin{equation}
\left(L_{-1}^2-g L_{-2}\right) \Phi_{(1| 2)}=0\;, \qquad \left(L_{-1}^2-\frac{1}{g} L_{-2}\right) \Phi_{(2|1)}=0\;.
\label{vir_nvc}
\end{equation}
The null vector conditions characterizing a field $\Phi_{(r|s)}$ yields  a differential equation of order $r s$ which is satisfied by any  conformal block containing $\Phi_{(r|s)}$. In particular, for the fields $\Phi_{(1|2)}$ and $\Phi_{(2|1)}$, this gives an order 2 differential equation which can be related to the Calogero-Sutherland Hamiltonian \cite{Cardy,Cardy_Doyon,ES,EBS}. 

Consider the most generic conformal block containing the field $\Phi_{(1|2)}$, namely
\begin{align}
 \langle\Phi_{(1|2)}(z_1) \Phi_{\Delta_2}(z_2)\dots \Phi_{\Delta_N}(z_N)\rangle \; .
\end{align}
Using standard contour deformation manipulations \cite{diFrancesco}, the null-vector condition \eqref{vir_nvc} can be cast in the differential form
\begin{align}
\mathcal{O}^{g}(z)\langle\Phi_{(1|2)}(z) \Phi_{\Delta_1}(z_1)\dots \Phi_{\Delta_N}(z_N)\rangle = 0 \label{12_diff_eq}
\end{align}
where the order 2 differential operator $\mathcal{O}^g(z)$ is 
\begin{equation}
\mathcal{O}^{g}(z) = \frac{\partial^2}{\partial z^2}-g\left(\sum_{j =1}^N\frac{\Delta_i}{(z-z_j)^2}+ \frac{1}{z-z_j}\frac{\partial}{\partial z_j} \right) \;. \label{Og}
\end{equation}
Likewise, conformal blocks containing the dual field $\Phi_{(2|1)}$ obey a similar differential equation, which can be obtained by simply changing $g \to 1/g$. 

\subsection{Heisenberg algebra $\textrm{H}$}
The CFT based on the Heisenberg algebra $\textrm{H}$ has  an additional $u(1)$ symmetry generated by a conserved current $J(z)$ of conformal dimension one.
As usual one defines the operator $a_n$ through the Laurent series of the  $J(z)$ current:
\begin{equation}
J(z) \Phi(w)\equiv \sum_{n}\frac{1}{(z-w)^{n+1}}a_{n} \Phi(w)  
\end{equation}
and they obey the so called Heisenberg algebra:
\begin{equation}
 [a_n,a_m] = n\delta_{n+m,0}\;. \\
\label{Heis}
\end{equation}
The stress energy tensor $T(z)$ of the theory is given by:
\begin{equation}
T(z)=\frac{1}{2}: J(z) J(z):
\end{equation}
where $::$ stands for the regularized product, and has central charge $c=1$. 
 The correspondent Virasoro operators $l_n$\footnote{Throughout the paper we use the notation $l_n$ to refer to the Virasoro operator associated to 
the $u(1)$ CFT.} are written in terms of $a_n$ as
\begin{eqnarray}
& & l_n =  \frac{1}{2}\sum_{m \in \mathbb{Z}} a_{n-m}a_m \qquad n \neq0 \\
& & l_0 = \sum_{m >0} a_{-m} a_m + \frac{1}{2}a_0^2\;.
 \label{vir_op_a}
 \end{eqnarray}
The $l_n$  commute with the $a_n$ in the following way
\begin{equation}
 [l_n,a_m] = - m a_{n+m}
 \end{equation}
 and form a Virasoro algebra with central charge $c=1$
\begin{equation}
[l_n,l_m] = (n-m)l_{n+m} + \frac{1}{12}n(n^2-1)\delta_{n+m,0}\;.
\label{viras_u1}
\end{equation}
  
  The simplest way to realize the $c=1$ theory is by introducing a free boson $\phi(z)$ normalized to
  \begin{equation}
\langle \phi(z) \phi(w) \rangle=-\ln (z-w)\;.
\end{equation}
In terms of this boson the current $J(z)$ reads
\begin{equation}
J(z)=i\partial \phi(z)\;.
\end{equation}   
The primaries  of the (\ref{Heis}) algebra are the vertex operators $V_{\beta} = :e^{i\beta \phi(z)}$:
\begin{eqnarray}
& & a_n V_{\beta}= 0\;,  \qquad n > 0 \\
& & a_0 V_{\beta} = \beta \,V_{\beta}
 \end{eqnarray}
where $\beta$ is the $U(1)$ charge. From the (\ref{vir_op_a}), it is easy to derive the conformal dimension $\Delta_{\beta}$ of the vertex $V_{\beta}$:
\begin{equation}
\Delta_{\beta}=\frac{\beta^2}{2}\;.
\end{equation}

From a vertex operator $V_{\beta}$ all possible independent descendant can be obtained by applying the $a_n$ operator, $V_{\beta}^{(n_1,\dots ,n_k)}=a_{n_1}\dots a_{n_k}V_{\beta}$ with $n_1\geq n_2\dots \geq n_k$. 
Note that, for the $c=1$ theory,  there are no singular vectors in this basis. Moreover the conformal block of $N$ vertex operator are easily computed,
\begin{equation}
\langle V_{\beta_1}(z_1)\dots V_{\beta_2}(z_N)\rangle=\prod_{i<j} z_{ij}^{\beta_i \beta_j} \quad \mbox{for}\quad \;\sum_{i}\beta_i=0\;.
 \end{equation}

\newsec{Virasoro models: separation of variables and duality of partitions }
\label{sec:MM}

In this section we introduce a set of Calogero-Sutherland eigenfunctions using conformal blocks in $\textrm{u}(1) \otimes \textrm{WA}_{k-1}(g)$, and we study their properties. Since the generic formulas for the $\textrm{WA}_{k-1}$ algebras are rather complicated, we start with the presentation of the Virasoro case ($k=2$) and postpone the treatment of  the generic case to section \ref{WA_section}.

\subsection{Separation of variables}

As explained in section \ref{CFT_section}, the fields  $ \Phi_{(1|2)}$ and $ \Phi_{(2|1)}$ are of special importance, as any conformal block involving them obey a second order differential equation \eqref{Og}. In order to build eigenfunctions of the Calogero-Sutherland Hamiltonian, we are led to consider functions of the form:
\begin{equation}
\langle \Phi_{(2|1)}(w_1) \cdots \Phi_{(2|1)}(w_M)\Phi_{(1|2)}(z_1) \cdots \Phi_{(1|2)}(z_N)\rangle_{a,b}  \label{conf_block}
 \end{equation}
where the various conformal blocks are labelled by the double index $a,b$. It is natural to use a double index because both the $M$ fields $\Phi_{(2|1)}$ and the $N$ fields $\Phi_{(1|2)}$ must fuse to the identity sector. There are $2^{N/2-1}$ such Bratelli diagrams for the fusions of $\Phi_{(1|2)}$, and $2^{M/2-1}$ for $\Phi_{(2|1)}$.

Upon multiplying by the correct $u(1)$ factors, this function was shown in \cite{EBS} to obey a differential equation involving two CS Hamiltonians, and exhibits separation of variables in the sense of \eqref{separation_variables}. We introduce the function $\mathcal{F}^{a,b}_{M,N}$
\begin{equation}
 \mathcal{F}^{a,b}_{M,N}(w;z)\equiv\langle \Phi_{(2|1)}(w_1) \cdots \Phi_{(2|1)}(w_M)\Phi_{(1|2)}(z_1) \cdots \Phi_{(1|2)}(z_N)\rangle_{a,b} \prod_{1\leq i <j}^{M}w_{ij}^{2\tilde h} \prod_{i,j}(w_i-z_j)^{1/2} \prod_{1\leq i<j}^{N}z_{ij}^{2h} \, ,
 \label{qhMM}
 \end{equation}
where we  denote for simplicity 
\begin{equation}
h=\Delta_{(1|2)}=\frac{3g}{4}-\frac{1}{2}\;, \qquad \tilde h=\Delta_{(2|1)}=\frac{3}{4g}-\frac{1}{2}\;.
\label{conf_dim_g}
\end{equation}
The $u(1)$ factors are such that the conformal block $\mathcal{F}^{a,b}_{M,N}(w;z)$ is regular whenever two fields come close to each other, as long as they have the following fusion channels
\begin{eqnarray}
\Phi_{(1|2)}(z_i) \Phi_{(1|2)}(z_j)&\sim&\frac{\mbox{I}}{(z_i-z_j)^{2h}} \label{pp_ope_MM}\;, \\
\Phi_{(2|1)}(w_i) \Phi_{(2|1)}(w_j) &\sim& \frac{\mbox{I}}{(w_i-w_j)^{2\tilde h}}\;  \label{ss_ope_MM}\;, \\
\Phi_{(1|2)}(z_i) \Phi_{(2|1)}(w_j)&\sim&\frac{\Phi_{22}}{(z_i-w_j)^{1/2}} \label{sp_ope_MM}\;.
\end{eqnarray} 
In the generic case $\Phi_{(1|2)} \times \Phi_{(1|2)}\to \mbox{I}$ is only one of the two possible fusion channels. However for $g = (r+2)/3$, the other fusion $\Phi_{(1|2)} \times \Phi_{(1|2)}\to \Phi_{(1|3)}$ vanishes. These are the cases of interest for the construction of trial wavefunctions using Jack polynomials in the fractional quantum Hall effect \cite{BernevigHaldane1,BernevigHaldane2,BernevigHaldane3,BernevigW}.
There the $u(1)$ factor $\prod_{i<j} (z_i-z_j)^{2h}$ is necessary to make the wave-function single valued in terms of the positions of the electrons $z_i$ \cite{MooreRead,ReadRezayi}.

As was shown in \cite{EBS}, the null-vector conditions (\ref{12_diff_eq}) for $\Phi_{(1|2)}$ and $\Phi_{(2|1)}$ can be combined together to obtain the following differential equation 
\begin{eqnarray}
\left[ h^{\alpha}(z)+g\;h^{\tilde \alpha}(w)\right]\mathcal{F}^{a,b}_{M,N}(w;z)=0
\label{dec_eq_MM}
\end{eqnarray}
where $h^{\alpha}$ belongs to the tower of commuting Calogero-Sutherland Hamiltonians
\begin{eqnarray}
 h^{\alpha}(z)\equiv \mathcal{H}^{\alpha}(z)-\E^{\alpha}_0+\(\frac{N-2}{ \alpha}-1\) [\mathcal{P}(z)-\mathcal{P}_0]-\frac{NM(M-2)}{4}\;,\\
  h^{\tilde \alpha}(w)\equiv \mathcal{H}^{\tilde\alpha}(w)-\E^{\tilde\alpha}_0+\(\frac{M-2}{ \tilde\alpha}-1\) [\mathcal{P}(w)-\mathcal{P}'_0]-\frac{NM(N-2)}{4}\;,
\label{dec_eq_MMbis}
\end{eqnarray}
with $\CP(z)=\sum z_i\partial_i$ and  $\mathcal{H}^{\alpha}$ given by \eqref{CS_g_g} with coupling
\begin{align}
\alpha=\frac{1}{1-g}& & \tilde \alpha= \frac{1}{1-g^{-1}} & & \alpha+ \tilde \alpha =1\;.
\label{alpha_g}
\end{align}
The constants $\E^{\alpha}_0\equiv \E^{\alpha}_{\lambda^0}$ and $\mathcal{P}_0\equiv \mathcal{P}_{\lambda^0}$ are given by
\begin{equation}
\E^{\alpha}_0=\frac{h}{3}N(N-2)(N(2g-1)-5g+4)\;, \qquad \mathcal{P}_0=N(N-2) h\;, \label{constants}
\end{equation}
while $\E^{\tilde \alpha}_0$ and $\mathcal{P}'_0$ are given by similar expressions with $g\to g^{-1}$ and $N\to M$ and $\lambda^0\to \lambda'^0$. 
The degree of homogeneity of  $\mathcal{F}^{a,b}_{M,N}(w;z)$ in both the variables $w$ and $z$ is 
\begin{equation}
\mathcal{P}(z)+\mathcal{P}(w)=N(N-2) h+ M(M-2)\tilde h+MN/2 \;
\end{equation}
and it is clear that generically this function cannot be expanded in polynomial eigenbases neither in $w$ nor in $z$. However, it can be expanded on non-polynomial 
eigenfunctions of $\mathcal{H}^{\alpha}(z)$ and $\mathcal{H}^{\tilde \alpha}(w)$ and  a duality property similar to that of section (\ref{sec:dual_sep_Ising}) holds
\begin{equation}
\label{dual_exp_MM}
\mathcal{F}^{a,b}_{M,N}(w;z)=\sum_\lambda \mathcal{F}_{\lambda'}^{\tilde \alpha,a}(w)\,\mathcal{F}_{\lambda}^{\alpha,b}(z)\;.
\end{equation}
This looks like the duality property \cite{Stanley, Macdonald, Gaudin} of the Calogero-Sutherland model $g\to 1/g$, with some differences. One difference consists in the boundary condition of the CS eigenfunctions, which force us to choose $\alpha=1/(1-g)$. The second is the non-abelian monodromy of the
conformal blocks which implies the non-polynomial nature of the eigenfunctions.

\subsection{Duality for the partitions}

\begin{figure}[h]
      \begin{center}
      \includegraphics[scale=.4]{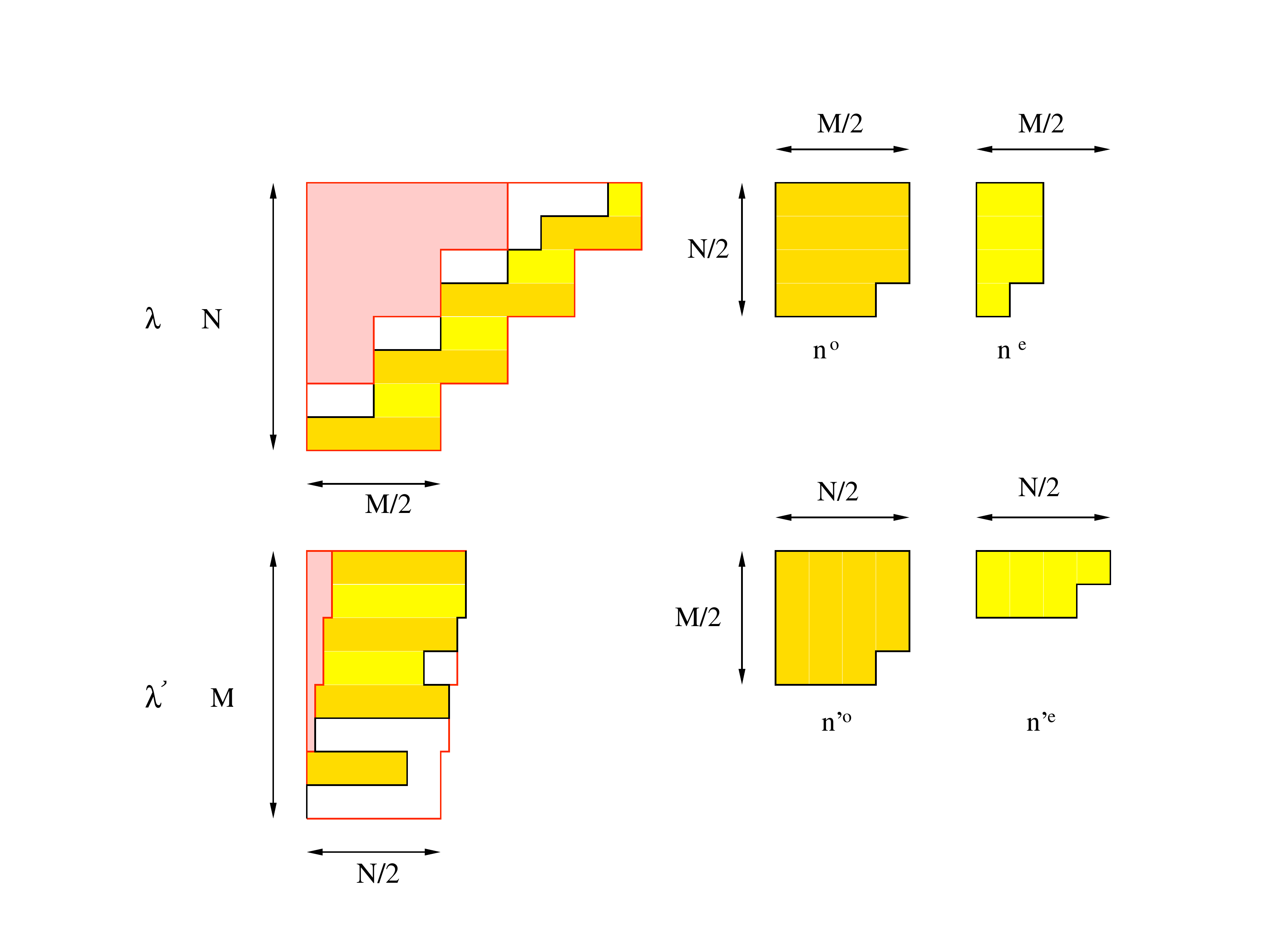}
      \end{center}
    \caption{ The relation between the sets of numbers $\lambda$ and $\lambda'$ is realized using two sets of dual Young diagrams, $n^o$,  $n'^o$
    and $n^e$,  $n'^e$. The blocks in pink correspond to $\lambda^0$ and  $\lambda'^0$, while the maximum envelopes, in red, correspond to
    $\Lambda^0\equiv M/2+\lambda^0$ and $\Lambda'^0\equiv N/2+\lambda'^0$. The maximum value of the partitions $n^o$ and $n^e$ is $(M/2)^{N/2}$.}
   \label{tableau}  
\end{figure}

Although neither $\mathcal{F}_{\lambda}^{\alpha,b}(z)$  nor $\mathcal{F}_{\lambda'}^{\tilde \alpha,a}(w)$ are polynomials, they are characterized by a ``partition"\footnote{We use the quotes in "partition" to stress that the parts $\lambda_i$ are generally not integers, which is related to the fact that the corresponding eigenfunctions are not polynomials. The dual ``partition'' $\lambda'$ is not the transpose of the ``partition'' $\lambda'$; they are related as specified by the equations (\ref{two_young}) and (\ref{two_young_dual}), see also Figure \ref{tableau}.} such that the CS eigenvalues are of the form  \eqref{CS_eig_for} for the couplings $\alpha$ and $\tilde{\alpha}$ as in \eqref{cs_gamma}
 \begin{align}
\E^{\alpha}_{\lambda} =\sum_{i=1}^N \lambda_i [\lambda_i+ \frac{1}{\alpha}(N+1-2i)]  \, ,& & 
\E_{\lambda'}^{\tilde{\alpha}} =\sum_{j=1}^M  \lambda'_j [\lambda'_j+\frac{1}{\tilde{\alpha}}(M+1-2j)]\, .
 \label{dual_energy}
  \end{align}
These ``partitions'' can be obtained by computing the behavior of $\CF_{\lambda}^{\alpha,b}(z)$ as  $z_1\gg z_2 \gg \dots\gg z_N$:
\begin{equation}
\CF_{\lambda}^{\alpha,b}(z_1,\dots,z_N) \sim z_1^{\lambda_1}  z_2^{\lambda_2} \cdots z_N^{\lambda_N} 
\end{equation}
and likewise as  $w_1\gg w_2 \gg \dots\gg w_M$ for $\CF_{\lambda'}^{\tilde \alpha,a}(w)$.  The result depends on the conformal block $b$ under consideration. Since the conformal blocks form a representation of the braid group and share the same eigenvalue, it is sufficient to treat a particular conformal block. We choose the first conformal block,  where the successive fields are fused two by two into the identity. 
In particular for $M=0$, the first conformal block
\begin{align}
\mathcal{F}^{1}_{0,N}(z) = \langle \Phi_{(1|2)}(z_1) \cdots \Phi_{(1|2)}(z_N)\rangle_{1} \prod_{1\leq i<j}^{N}z_{ij}^{2h} 
\end{align}
behaves in the limit $z_1\gg z_2 \gg \dots\gg z_N$ as
\begin{equation}
z_1^{2h(N-2)}z_2^{2h(N-2)}z_3^{2h(N-4)}z_4^{2h(N-4)}\dots\;z_{N-1}^0z_{N}^0\;
\end{equation}
and we encode this in a ``partition'' $\lambda^{0}$
\begin{align}
\lambda^{0}_{2i-1}=\lambda^{0}_{2i} =2h(N-2i)\;, & &  i=1,\dots,N/2\;.
\label{def_l_o}
\end{align}
One can already check that this ``partition'', when plugged in \eqref{CS_eig_for}, is consistent with the CS eigenvalue $\E^{\alpha}_0$  and the degree $\mathcal{P}_0$ \eqref{constants}.  Moreover in the polynomial case $2h$ is an integer and $\lambda_0$ becomes a true partition (with integer parts). This happens for $g=(r+2)/3$: we recover the densest $(k=2,r)$ admissible partition \cite{FJMM,FJMM2} and the corresponding eigenfunction is a Jack polynomial.
Similarly for $\mathcal{F}^{1}_{M,0}(w)$, we infer that the ``partition'' corresponding to the lowest eigenstate in $w$ is given by
     \begin{align}
 \lambda'^{0}_{2j-1}= \lambda'^{0}_{2j}=2\tilde h(M-2j)\;, & & j=1,\dots,M/2\;.
  \end{align}
  and it corresponds under the duality to the maximum ``partition''
    \begin{eqnarray}
      \label{def_lambda_max_g}
  \Lambda^0_i&=&\lambda^0_i+\frac{M}{2}\;.    
  \end{eqnarray}
We now turn to the description of the  excited states $\lambda$ appearing in the expansion (\ref{dual_exp_MM}). It is likely that they differ from the ground state $\lambda^0$ by $\lambda_i-\lambda^0_i=n_i$ with $n_i$ positive integers, since they can be constructed from the ground state by applying some "creation operators" \cite{LPV} which increase degree of homogeneity by one. Moreover the relation \eqref{dec_eq_MM} implies a relationship between the ``partitions'' which characterize the eigenfunctions of the two Hamiltonians $\mathcal{H}^{\tilde{\alpha}}(w)$ and $\mathcal{H}^{\alpha}(z)$. 
We conjecture that the sets $\lambda$ and $\lambda'$ are characterized by two sets of dual partitions $n^{e,o}$ and $n'^{e,o}$ in the following manner (see Fig. \ref{tableau})
      \begin{eqnarray}
      \label{two_young}
\lambda_{2i-1}=\Lambda^{0}_{2i-1}-n^e_{N/2-i+1}\;, &\quad & \lambda_{2i}=\Lambda^{0}_{2i}-n^o_{N/2-i+1}\;,\\ 
\lambda'_{2i-1}=\lambda'^{0}_{2j-1}+n'^o_{j}\;, &\quad& \lambda'_{2j}=\lambda^{0}_{2j}+n'^e_{j}\;,
 \label{two_young_dual}\\ \label{two_constraints}
 {\rm with} \quad  n_i^{e,o}\leq M/2\quad &{\rm and}& n'^{e,o}_i\leq N/2\;.
  \end{eqnarray}
For a partition $n$ with lines of length $n_i$ the dual partition $n'$, with lines of length $n'_j$, is the partition where the lines of $n$ become the columns of $n'$. For two partitions $n$ and $n'$ dual to each other the following relations hold \cite{Stanley}
   \begin{eqnarray}
     \label{bstan}
b(n)\equiv 2\sum_i (i-1)n_i=\sum_j {n'_j(n'_j-1)}\;,\qquad |n|\equiv \sum_i n_i=|n'|\;.
 \end{eqnarray}

We can check this conjecture by evaluating the eigenvalues of $\mathcal{H}^{\alpha}(z)$ and $\mathcal{H}^{\tilde \alpha}(w)$ on the corresponding states, eigenvalues which are given in equation (\ref{CS_eig_for}) 
\begin{align}
\E^{\alpha}_{\lambda}  = \sum_{i=1}^N \lambda_i [\lambda_i+(1-g)(N+1-2i)] \, , & & 
\E_{\lambda'}^{\tilde \alpha}  = \sum_{j=1}^M  \lambda'_j [\lambda'_j+(1-g^{-1})(M+1-2j)]\;.
 \label{dual_energy-MM}
  \end{align}
Substituting the expressions (\ref{two_young}), (\ref{two_young_dual}) in the above formulas and using the notations from equation (\ref{bstan}) we obtain now the  expressions of the energies of the intermediate states purely in terms of the partitions $n^{e,o}$ and $n'^{e,o}$ as
  \begin{eqnarray}
\E_\lambda^{\alpha}
\label{en_deux_part_MM}
=\[b(n'^o)+b(n'^e)\]-g\[b(n^o)+b(n^e)\]+((1-g)N-M+g)(|n^o|+|n^e|)+2(g-1)|n^e|+\E_{\Lambda^0}^{\alpha}\;
 \end{eqnarray} 
 and 
   \begin{eqnarray}
\E_{\lambda'}^{\tilde \alpha}
=\[b(n^o)+b(n^e)\]-\frac{1}{g}\[b(n'^o)+b(n'^e)\]+\frac{(2-g)M+2g-3}{g}(|n'^e|+|n'^o|)+\frac{2(g-1)}{g}|n'^o|+\E_0^{\tilde \alpha}\;.
 \end{eqnarray} 
It is a non-trivial check of the conjectured expressions (\ref{two_young}), (\ref{two_young_dual}) that the two eigenvalues $\E_{\lambda}^{\alpha}$ and  $\E_{\lambda'}^{\tilde \alpha}$ satisfy the duality condition implied by (\ref{dec_eq_MM})
\begin{eqnarray} 
\nonumber
\E_\lambda^{\alpha}-\E_0^{\alpha}+g(\E_{\lambda'}^{\tilde \alpha}-\E_{0}^{\tilde \alpha})+\(\frac{N-2}{\alpha}-1\)\({\cal P}_\lambda-{\cal P}_0\)+g\(\frac{M-2}{\tilde \alpha}-1\)\({\cal P}_{\lambda'}-{\cal P}_{0'}\)=\\
\E_{\Lambda^0}^{\alpha}-\E_0^{\alpha}+\frac{NM}{2}\(\frac{N-2}{ \alpha}-1\)=\frac{NM(M-2)}{4}+\frac{gMN(N-2)}{4}\;.
 \end{eqnarray} 
Although the manipulations from this section may seem too abstract, due in particular to the fact that the properties of the non-polynomial eigenfunctions of 
the Calogero-Sutherland model are largely unexplored, one can stick to the particular case $g=4/3$, which corresponds to the Ising CFT. In this case, the dimension of the fermion is $1/2$, so that $2h=1$ and the ``partition'' $\lambda_0$ becomes a true partition, and the associate eigenfunctions
indexed by $\lambda$ are Jack polynomials with clustering properties. This case is well under control, and we treat this particular case in the Appendix \ref{app: Ising_explicit}.
As an example, we give the explicit expressions of the non-polynomial eigenfunctions $\CF^{4,a}_{\lambda'}(w)$ for $M=4$ and $N=2$.
In the Ising case, due to the constraint of $(2,2,N)$ admissibility of the partition $\lambda$, (\ref{admissible_part}),  $n^o$ and $n^e$  obey the extra mutual constraints
   \begin{eqnarray}
 0\leq n^e_i\leq n^o_i\leq M/2\;,  \qquad  n_{i+1}^o\leq n_i^e+2\;.
  \end{eqnarray}
  It is interesting to know whether in general there are any constraints for the partitions $n^e$ and $n^o$ in addition to the ones in equation 
  (\ref{two_constraints}).
  
 Let us comment on the significance of the expressions (\ref{two_young}) and (\ref{two_young_dual}).
 Remembering that the energy for a polynomial eigenfunction of the Calogero-Sutherland model with $N$ particles indexed by the partition $n$ can be written as
    \begin{eqnarray}
     \label{bstanE}
\CE_n^\alpha=\sum_{i}^N n_i\left[n_i+\frac{1}{\alpha}(N+1-2i)\right]=b(n')-\frac{1}{\alpha}b(n)+\(\frac{N-1}{\alpha}+1\)|n|\;.
 \end{eqnarray} 
the expression (\ref{two_young}) suggests that  the intermediate states $\lambda$ are described {\bf by two} Calogero-Sutherland models, each with $N/2$ particles, with eigenfunction indexed by the two partitions $n^e$ and $n^o$ and at coupling constant $\alpha=1/g$. 
 States which are indexed by a pair of Young diagrams appeared in the expression of the Nekrasov's instanton partition function  \cite{Nekrasov} which can be related to the conformal blocks of the Liouville theory \cite{AGT}. Since Liouville theory can be treated exactly in the same fashion as the generic Virasoro models in this section, provided that we change the sign of the coupling constant $g$, one can suspect that the basis we have identified in this section is related to the basis used to prove the AGT conjecture \cite{AGT,AFLT}.
In the next section we are going to show that this basis indeed corresponds to the basis considered by Alba, Fateev, Litvinov and Tanopolsky
in \cite{AFLT}.

\section{Hidden integrable structure in $u(1) \otimes \textrm{Virasoro}$  modules}
\label{sec_uone_MM}

\subsection{The $u(1)$ sector and the Coulomb gas representation of the minimal model}

An essential feature in proving the duality (\ref{dec_eq_MM})  is the presence of the term $\prod_{i,j}(z_i-w_j)^{1/2}$ which dresses the conformal block 
in $\mathcal{F}^{a,b}_{M,N}(w;z)$. This factor can be accounted for by introducing an $u(1)$ component in the CFT. In the context of the quantum Hall effect, the
$u(1)$ component carries the electric charge. The corresponding conserved current is $J(z)=i\partial \phi(z)$. The $u(1)$ factors introduced in the previous section where inherited from the fractional quantum Hall effect, and it turns out to be more convenient to work with a slightly modified function
 \begin{eqnarray} 
{F}^{a,b}_{M,N}(w;z)\equiv \mathcal{F}^{a,b}_{M,N}(w;z)\prod_{1\leq i< j}^N z_{ij}^{1-g} \prod_{1\leq i< j}^M w_{ij}^{1-g^{-1}}\;.
 \end{eqnarray} 
 This slight modification does not spoil the separation of variables since it does not mix the variables $z$ and $w$. However it changes the Laplace-Beltrami operators $\mathcal{H}^{\alpha}$  and $\mathcal{H}^{\tilde{\alpha}}$  involved in the differential equation  \eqref{dec_eq_MM}-\eqref{dec_eq_MMbis} to the proper Calogero-Sutherland Hamiltonian $H_2^g$ and $H_2^{1/g}$, respectively.
  
 The modified function $F^{a,b}_{M,N}(w;z)$ corresponds to conformal blocks of the product theory $u(1) \otimes \textrm{Vir}(g)$
 \begin{align} 
{F}^{a,b}_{M,N}(w;z)=\langle \tilde V (w_1)\ldots \tilde V (w_M)\,V (z_1)\ldots V(z_N)\rangle_{a,b} \;, \label{Corr_V}
 \end{align} 
 where the fields $V$ and $\tilde{V}$ are 
  \begin{eqnarray} 
  \label{vaPhi}
V (z)\equiv \Phi_{(1|2)}(z) \,e^{i\sqrt{\frac{g}{2}}\phi(z)}  \;, \qquad   \tilde V (w)\equiv \Phi_{(2|1)}(w) \,e^{i\frac{1}{\sqrt{2g}}\phi(w)} \;.
 \end{eqnarray}
 The minimal model can be represented by a Coulomb Gas \cite{DotsenkoFateev} construction, \emph{i.e.} with the help of a bosonic field $\varphi(z)$, with some constraints which include the null vector conditions
 (\ref{vir_nvc}). Denoting the Fourier modes of the bosonic field $\varphi(z)$ by $b_n$, the generators of the Virasoro algebra (\ref{viras}) are given by
 \begin{eqnarray}
L_n =  \frac{1}{2}\sum_{m \in \mathbb{Z}} :b_{n-m}b_m: -\alpha_0(n+1)b_n\;, \quad {\rm with}\quad 2\alpha_0=\sqrt{\frac{2}{g}}-\sqrt{2g}\;.
 \label{vir_op_b}
 \end{eqnarray}
 The operator with conformal dimension $\Delta_\alpha=\alpha(\alpha-2\alpha_0)/2$ can be represented with one of the vertex operators
  \begin{eqnarray}
\Phi_{\alpha}(z)\sim :e^{i\alpha\varphi(z)}:\; \quad {\rm or}\quad  \Phi_{\alpha}(z)\sim :e^{i(2\alpha_0-\alpha)\varphi(z)}:\;.
 \label{vert_op_b}
 \end{eqnarray}
 The identification of the vertex operators with reflected charge $\alpha$ and $2\alpha_0-\alpha$ is a non-trivial property of the theory. 
 Let us consider the first possibility for the field identification in (\ref{vert_op_b}); we can then write  for the fields in (\ref{vaPhi})
  \begin{eqnarray} 
  \label{va_vb}
V (z)=:e^{i\sqrt{\frac{g}{2}}[\phi(z)+\varphi(z)]} : \;, \qquad   \tilde V (w)=:e^{i\frac{1}{\sqrt{2g}}[\phi(w)-\varphi(z)]} :\;.
 \end{eqnarray}
 Let us note that the Virasoro modes  $L_n$ in (\ref{vir_op_b}) are left invariant under simultaneous change of sign of the bosonic modes  $b_n$ and of the charge at infinity $\alpha_0$, which in turn is equivalent to $g\to 1/g$. This operation exchanges the operators $V(z)$ and $\tilde V(z)$ in equation (\ref{va_vb}) and corresponds to the duality transformation. It is tempting to relate the decoupling relation (\ref{dec_eq_MM}) to the fact that the two vertex operators are built from commuting bosonic fields.
 
\subsection{The Hilbert space and the Calogero-Sutherland integrals of motion}
\label{corr}

The separation of variables  \eqref{dec_eq_MM} and \eqref{dual_exp_MM} can be seen as a corollary of a deeper connection between Calogero-Sutherland Hamiltonians and 
integrals of motions acting in the modules of $Vir (g)\otimes \mathcal{H}$. The idea is to interpret the decomposition 
 \begin{align} 
\langle \tilde V (w_1)\ldots \tilde V (w_M)\,V (z_1)\ldots V(z_N)\rangle_{a,b} = \sum_\lambda F_{\lambda'}^{1/g,a}(w) \, F_{\lambda}^{g,b}(z) \label{separation2}
 \end{align} 
as coming from the insertion of a complete basis of descendants between the operators $\tilde{V}(w)$ and $V(z)$, with a very particular basis ensuring that each function $F_{\lambda}^{g,b}(z)$ is an eigenvector of the CS Hamiltonians $H^n_{g}$. Such a basis is in fact unique, and one of the main result of this paper is the construction of such a basis. 
In this section we present this result, while more details about its derivation can be found in Appendix \ref{Correspondence}.  

We focus on the fields $V$ from \eqref{vaPhi}, the results for $\tilde{V}$ being simply obtained by substituting $g \to 1/g$. We are therefore concerned with correlation functions of the form
\begin{align}
f^{+}_{\mu} (z_1,z_2,\cdots,z_N) &  = \langle \mu | V(z_1) V(z_2) \cdots V(z_N) | P \rangle \label{fap} \\
f^{-}_{\mu} (z_1,z_2,\cdots,z_N) &  = \langle P | V(z_1) V(z_2) \cdots V(z_N) | \mu \rangle
\end{align}
where $| \mu \rangle$ is an arbitrary field (primary or descendant), and $|P\rangle$ is a primary field (i.e. annihilated by $a_m$ and $L_m$ for $m>0$) and we dropped the conformal block label as it does not play a role in this analysis.  The order $n$ CS Hamiltonian $H^g_{r}$ acting on the variables $(z_1,\cdots,z_N)$ can be rewritten as an operator $I^{\pm}_{r+1}$ acting on the state $\mu$\footnote{The basis $I^\pm_{r+1}(g)$ we will use in the following corresponds in fact to a combination of $H_{s}^g$ with $s\leq r$, see {\it e.g.} (\ref{const_MM}).}:
\begin{equation}
H_{r}^g f^{\pm}_{\mu} (z_1,z_2,\cdots,z_N)  =  \sum_{\nu} \left[I^{\pm}_{r+1}(g)\right]_{\mu,\nu} f^{\pm}_{\nu} (z_1,z_2,\cdots,z_N) \label{Cor}
\end{equation}
To put it differently, $f^{\pm}_\mu(z_1,\cdots,z_N)$ is an eigenstate of $H_r(g)$ iff $|\mu\rangle$ is an eigenstate of $I_{r+1}^{(\pm)}(g)$.
We checked this correspondence and computed the value of $I_{r+1}^{\pm}(g)$ for $r=2,3$ in Appendix \ref{Correspondence}.

This expression for $H_2^g$ comes from the degeneracy at level $2$ in the module of the operator $V$:
\begin{equation}
\left(L_{-1}^2 - g L_{-2} \right)V=0
\end{equation}
By standard contour deformation (see Appendix \ref{Correspondence} for more details), this relation yields a differential equation of order 2 for any correlator involving $V(z)$. For a symmetric correlation function of the form \eqref{fap}, this differential equation becomes the order 2 CS Hamiltonian 
\begin{equation}
H_2^g=\sum_{i=1}^N \left(z_i \frac{\partial}{\partial {z_i}}\right)^2 + g(1-g)\sum_{i\neq j} \frac{z_iz_j}{z_{ij}^2}
\end{equation} 
up to an extra term corresponding to the contours being at infinity, yielding an operator acting on $\langle \mu|$. This is of course the operator $I_3^+(g)$.
The operator which acts on $| \mu\rangle$ is $I_3^-(g)$ with
\begin{equation}
I_3^{(\pm)}(g) = 2(1-g)\sum_{m\geq 1} m a_{-m}a_m  \pm \sqrt{2g} \sum_{m\neq 0} a_{-m} L_{m} \pm  \sqrt{\frac{g}{2}}\left(  \sum_{m,k \geq 1} a_{-m-k}a_m a_k  +  a_{-m}a_{-k}a_{m+k} \right)\;.
\label{I3vir}
\end{equation}
The relation for $H_3^g$ can be obtained from the degeneracy at level 3
\begin{equation}
(L_{-1} +3\sqrt{g/2}a_{-1})\left(L_{-1}^2 - g L_{-2} \right)V=0
\end{equation}
and the explicit expression for $I_4^{\pm}(g)$ is given in (\ref{Iquatre}). $I^{-}_4(g)$ coincides, up to the change $g\to -g$ and a change in normalizations of the bosonic operator, with the operator $I_4(g)$ which appeared in \cite{AFLT} in the context of the AGT conjecture.
We conjecture that this structure holds true for any $r$, in the sense that the higher integrals of motion would correspond to a particular descendent
of the second order degenerate vector
\begin{equation}
\sum_{n_1+\ldots+n_l=r-2}c_{n_1,\ldots,n_l}\,L_{-n_1}L_{-n_2}\ldots a_{-n_{l-1}}a_{-n_l}\left(L_{-1}^2 - g L_{-2} \right)V=0
\end{equation}
with coefficients $c_{n_1,\ldots,n_l}$ to be determined.
 This would define two towers of commuting integral of motions $I_{r}^+(g)$ and $I_{r}^-(g)$, charge conjugate from one another.   

In the module of the primary field $P$, \emph{i.e.} the set of all $Vir (g)\otimes \mathcal{H}$  descendants of $P$, one can diagonalize the operators $I_r^{(\pm)}(g)$. We denote the corresponding basis of descendants $|P_\lambda^{\pm}(g) \rangle$.  These two orthogonal bases  are charge conjugate from one another. In this basis, the OPE of $N$ vertex operator $V$ 
\begin{align}
V(z_1)\cdots V(z_N) |P \rangle & = \sum_{\lambda } F_{\lambda}^{(g,+)}(z_1,\cdots,z_N)| P_\lambda^+(g) \rangle \\
\langle P | V(z_1)\cdots V(z_N) & = \sum_{\lambda} F_{\lambda'}^{(g,-)}(z_1,\cdots,z_N) \langle  P^-_\lambda(g) | 
\end{align}
enjoys a natural action of Calogero-Sutherland Hamiltonians. All the $N$-point functions 
\begin{align}
F_{\lambda}^{(g,+)}(z_1,\cdots,z_N) = \langle  P_\lambda^+(g) | V(z_1)\cdots V(z_N) |P \rangle  \\ F_{\lambda'}^{(g,-)}(z_1,\cdots,z_N) = \langle  P | V(z_1)\cdots V(z_N) | P_\lambda^-(g) \rangle
\label{def_F} 
\end{align}
 are simultaneously eigenstates of the whole tower of Hamitlonians $H_{n}^g$ because of the  correspondence  \eqref{Cor}.

As the CFT remains unchanged under $g \to 1/g$, one could think of introducing another two bases, namely $|P_\lambda^{\pm}(1/g)\rangle$. However the operators $I_r^{\pm}(g)$ are self dual in the sense:
\begin{equation}
I_r^{(\pm)}(g) \propto I_{r}^{(\mp)}(1/g)
\end{equation}
and these bases are related through
\begin{equation}
|P_\lambda^{\pm}(g)\rangle = |P_\lambda^{\mp}(1/g)\rangle .
\end{equation}
This relation induces the duality \eqref{dec_eq_MM}, as the \eqref{Cor} implies the following structure for the $M$ points OPE  of $\tilde{V}$:
\begin{align}
\tilde{V}(w_1)\cdots \tilde{V}(w_M) |P \rangle & = \sum_{\lambda} F_{\lambda}^{(1/g,+)}(w_1,\cdots,w_M)|P_\lambda^{-}(1/g)\rangle
\end{align}
where  $ \tilde{F}_{\lambda}^{(1/g,+)}(z_1,\cdots,z_N)$ diagonalize all CS Hamiltonians $H_r^{1/g}$.  

Upon inserting a complete basis of descendants $|P_\lambda^{-}(g)\rangle = |P_\lambda^{+}(1/g)\rangle$ between the $V$'s and $\tilde{V}$'s in the mixed correlator, we obtain a generic separation of variables
\begin{equation}
\langle  P | V(z_1) V(z_2) \cdots V(z_N)  \tilde{V}(w_1) \cdots \tilde{V}(w_M) |P \rangle = \sum_{\lambda} F_{\lambda}^{(g,-)}(z_1,\cdots,z_N) F_{\lambda}^{(1/g,+)}(w_1,\cdots, w_M)
\end{equation} 
and we recover  \eqref{separation2} when we choose the primary $P $ to be the identity.

\subsection{The Virasoro model at $g=1$}
\label{minmodg1}

As noticed by Belavin and Belavin for the Liouville case \cite{BB}, at $g=1$ the structure of the Hilbert space of the conformal field theory and the duality become particularly transparent, in particular we can better understand the role of  the extra $u(1)$ component.
After a change of basis, the theory can be described with two copies
of independent bosons, coupled only by zero modes. We give the details of the construction below and we use the definitions from the section \ref{sec_uone_MM} for the bosonisation of the minimal model. Let us note that at $g=1$ the charge at infinity $\alpha_0$ defined in (\ref{vir_op_b}) vanishes and the stress-energy tensor is purely quadratic in the bosonic field.

The first non-trivial integral of motion, $I_3^{\pm}$ is in this case cubic in the bosonic fields,
\begin{equation}
\label{I3}
I_3^{\pm}(1)=\pm\[\sqrt{2}\sum_{m\neq 0}a_{-m}L_m+ \frac{1}{\sqrt{2}}\sum_{m,k>0}\(a_{-m-k}a_ma_k+a_{-m}a_{-k}a_{m+k}\)\]\;.
 \end{equation}
 Moreover, it is the odd in the operators $a_m$, so that
 \begin{equation}
a_m\to -a_m \quad {\rm sends}\quad I_3^{+}(1)\to I_3^{-}(1)=-I_3^{+}(1)
 \end{equation}
and  it is even in the operators $b_m$, since $L_m$ is quadratic in the $b_m$'s. The next integral of motion, $I_4^{\pm}$ is even in both sets of bosonic creation/annihilation operators
$a_m$ and $b_m$,
\begin{eqnarray}
I_4^{\pm}(1)&=&- \sum_{m >0 } L_{-m}L_{m}   
-\frac{3}{2} \sum_{m,p >0 } \left(  2L_{-p}a_{-m}a_{p+m} + 2 a_{-m-p}a_{m}L_{p} + a_{-m}a_{-p}L_{p+m} + L_{-m-p}a_{m}a_{p} \right)  \nonumber \\
&-&\frac{1}{2} L_0^2 - 3 L_0 \sum_{m>0} a_{-m}a_m 
 - \frac{1}{8} \sum_{\substack{m_1+m_2+m_3+m_4 = 0 \\ m_i \neq 0}} :a_{m_1}a_{m_2}a_{m_3}a_{m_4}:
-  \frac{1}{2} \sum_{m\geq 1}m^2  a_{-m}a_m
\end{eqnarray}
We are going to show that these two integrals of motion can be separated each into sums of two integrals of motion for independent bosons, plus a part containing the zero mode $b_0$. 
Let us rotate the bosonic basis and define the new bosonic operators
 \begin{equation}
c_m=\frac{1}{\sqrt{2}}\(a_m+b_m\)\;, \qquad \tilde c_m=\frac{1}{\sqrt{2}}\(a_m-b_m\)
 \end{equation}
and define the mutually commuting Hamiltonians $\CI_2(c)$, $\CI_3(c)$  and $\CI_4(c)$ as 
  \begin{eqnarray}
   \label{H_deux_one}
   \CI_2(c)=\sum_{m>0}c_{-m}c_m
    \end{eqnarray}
  \begin{eqnarray}
   \label{H_trois_one}
   \CI_3(c)=\sum_{m,k>0}\(c_{-m-k}c_mc_k+c_{-m}c_{-k}c_{m+k}\)
    \end{eqnarray}
     \begin{equation}
       \label{H_quatre_one}
\CI_4(c)=-\frac{1}{2}\sum_{m>0}m^2 c_{-m}c_m-\frac{1}{4} \sum_{\substack{m_1+m_2+m_3+m_4 = 0 \\ m_i \neq 0}} :c_{m_1}c_{m_2}c_{m_3}c_{m_4}:\;.
 \end{equation}
The Hamiltonian (\ref{H_trois_one}) is  known to be the one-component Calogero-Sutherland Hamiltonian at $g=1$ expressed in collective variables \cite{Jevicki,AMOS}  while (\ref{H_quatre_one}) is the next corresponding conserved charge. It is likely that there exists a whole tower of conserved charges $\CI_r(c)$,
each being of total degree $r$ in the bosonic operators.
Their joint eigenfunctions are given by the Schur polynomials, 
 \begin{equation}
|n\rangle=S_n(c)|0\rangle\;.
 \end{equation}
The eigenstates are indexed by partitions $n$ and the corresponding eigenvalues are given by the simple formulas
  \begin{equation}
    \label{E_deux}
e_{2,n}=\sum_i n_i=|n|\;,
 \end{equation}
    \begin{equation}
    \label{E_trois}
e_{3,n}=\sum_i n_i(n_i-2i+1)=b(n')-b(n)\;,
 \end{equation}
    \begin{equation}
e_{4,n}=-\sum_i \[\(n_i-i+\frac{1}{2}\)^3+\(i-\frac{1}{2}\)^3\]-\frac{1}{4}\sum_i n_i\;.
 \end{equation}
where $b(n)$ is defined in equation (\ref{bstan}). On the expression (\ref{E_trois}) it is obvious that dual partitions $n$ and $n'$ have opposite energies $e_{3,n}=-e_{3,n'}$. It is slightly more complicated to show that $e_{4,n}=e_{4,n'}$. On the Schur polynomials, the duality acts like $S_n(-c)\sim S_{n'}(c)$ where $n'$ is the partition dual to $n$. These findings are consistent with the fact that changing the sign of the bosonic operators $c_k$ changes the sign of the Hamiltonian $\CI_3(c)$
and it leaves  $\CI_4(c)$ invariant.
 
 The Hamiltonian $I_3(1)$ can be written as a sum of two commuting Hamiltonians depending on the bosonic modes $c_m$ and $\tilde c_m$
 \begin{eqnarray}
   \nonumber 
I_3^+(1)&=&\frac{1}{\sqrt{2}}\sum_{m,k>0}\(a_{-m-k}a_ma_k+a_{-m}a_{-k}a_{m+k}\)+\frac{1}{\sqrt{2}}\sum_{m\neq 0,k\in Z}a_{-m}b_{m-k}b_k\\
&=&\CI_3(c)+\CI_3(\tilde c)+\sqrt{2}b_0(\CI_2(c)-\CI_2(\tilde c))\;,
\label{hamsum}
 \end{eqnarray}
 the two copies being only related by the zero mode $b_0$. On the module of the identity, the last term in the previous expression vanishes. A similar property is valid for the next conserved charge,
 \begin{eqnarray}
I_4^+(1)=\CI_4(c)+\CI_4(\tilde c)-3{\sqrt{2}}{b_0}[\CI_3(c)-\CI_3(\tilde c)]-\frac{3b_0^2}{2}[\CI_2(c)+\CI_2(\tilde c)]-\frac{b_0^4}{8}\;.
\label{ham_quatre_sum}
 \end{eqnarray}
The immediate consequence of the separation (\ref{hamsum}, \ref{ham_quatre_sum}) is that the eigenfunctions are factorized, for example for the module of the identity
    \begin{equation}
    \label{schbasis}
|n^o,n^e\rangle=\[S_{n^o} (c) S_{n^e}(\tilde c)+S_{n^o} (\tilde c) S_{n^e}(c)\]|0\rangle
 \end{equation}
where $S_n (c)$ is the Schur polynomial associated to the partition $n$ constructed from the creation operators $c_{-k}$. We have isolated the combination which is symmetric in $b_k\to -b_k$, since this is what we get from the minimal model by constructing the descendants using $L_{-k}$. 
 The corresponding energy is the sum of the two independent energies
   \begin{equation}
    \label{schenergy}
E_{r;\,n^o,n^e}=e_{r,n^o}+e_{r,n^e}\;, \qquad r=3,4\;.
 \end{equation}
 and this agrees with the equation (\ref{en_deux_part_MM}).
 To illustrate the construction of the eigenstates, we give below the three eigenvectors at level 2 for $b_0=0$ obtained by direct diagonalization of $I_3$ and $I_4$,
  \begin{eqnarray}
  |\bl2\br,\bl\emptyset\br\rangle=\sqrt{2}\[a_{-2}+\frac{1}{\sqrt{2}}(a_{-1}^2+b_{-1}^2)\]|0\rangle  &=&  \[c_{-2}+c_{-1}^2+ c\to \tilde c\]|0\rangle\\
|\bl1,1\br,\bl\emptyset\br\rangle = \sqrt{2}\[-a_{-2}+\frac{1}{\sqrt{2}}(a_{-1}^2+b_{-1}^2)\]|0\rangle &=&\[-c_{-2}+c_{-1}^2+ c\to \tilde c\]|0\rangle\\
|\bl1\br,\bl1\br\rangle=2\[a_{-1}^2-b_{-1}^2\]|0\rangle&=&  c_{-1}\tilde c_{-1}|0\rangle\;.
 \end{eqnarray}
 Modulo the normalizations, these vectors coincide with the ones of Belavin and Belavin \cite{BB}.
 The corresponding eigenvalues of $I_3$ are  $2,-2,0$, while for $I_4$ they are $-4,-4,-1$. 
 
 For states which are not in the module of the identity but have a charge $q$,  $b_0|q\rangle=q|q\rangle$, we have
  \begin{equation}
|n^o,n'^e;q\rangle=S_{n^o} (c) S_{n^e}(\tilde c)|q\rangle+S_{n^o} (\tilde c) S_{n^e}(c)|-q\rangle\;.
 \end{equation}
 
 \subsection{Arbitrary $g$}
 \label{minmodggen}
 
 The separation of the energy of the intermediate states into two independent parts (\ref{en_deux_part_MM}) begs for an explanation. We have seen in 
 the previous section that  at $g=1$ this separation originates in the separation of the hamiltonian $I_3$ into two commuting parts. In this section
 we are going to investigate how the Hamiltonian $I_3$ can be written in terms of two independent bosons.
 First, we define the following one-component bosonic Calogero-Sutherland Hamiltonians \cite{AMOS,Jevicki}
   \begin{eqnarray}
   \label{H_deux_g}
   \CI_2(c)=\sum_{m>0}c_{-m}c_m\;,
    \end{eqnarray}
 \begin{eqnarray}
  \label{H_trois_g}
   \CI_3^\pm(c;g)=(1-g)\sum_{m>0}mc_{-m}c_m\pm \sqrt{g}\sum_{m,k>0}\(c_{-m-k}c_mc_k+c_{-m}c_{-k}c_{m+k}\)\;,
    \end{eqnarray}
     \begin{eqnarray}
       \label{H_quatre_g}
\CI_4^\pm(c;g)&=&\(\frac{3g}{2}-g^2-1\)\sum_{m>0}m^2 c_{-m}c_m-\frac{g}{4} \sum_{\substack{m_1+m_2+m_3+m_4 = 0 \\ m_i \neq 0}} :c_{m_1}c_{m_2}c_{m_3}c_{m_4}:
\pm\\ \nonumber
&\pm&3\sqrt{g}(g-1)\sum_{m,l>0}m(c_{-m-l}c_mc_l+c_{-m}c_{-l}c_{m+l})\;.
 \end{eqnarray}
 They obey the following symmetry properties
  \begin{eqnarray}
    \nonumber 
\CI_m^+(c;g)&=&\CI_m^-(-c;g)\;, \qquad m=3,4\;,\\ \label{E_trois_g_sym}
\CI^+_3(c;g)&=&-g\,\CI^-_3(c;1/g)=-g\,\CI^+_3(-c;1/g)\;,\\ \nonumber
 \quad\CI^+_4(c;g)&=&g^2\,\CI^-_4(c;1/g)=g^2\,\CI^+_4(-c;1/g)
 \end{eqnarray}
and their eigenvalues are given by 
  \begin{equation}
    \label{E_deux_g}
e_{2,n}(g)=\sum_i n_i=|n|\;,
 \end{equation}
  \begin{equation}
    \label{E_trois_g}
e^+_{3,n}(g)=\sum_in_i\[n_i-g(2i-1)\]=b(n')-gb(n)+(1-g)|n|=-g\,e^+_{3,n'}(1/g)\;,
 \end{equation}
    \begin{equation}
e^+_{4,n}(g)=-\sum_i \[\(n_i-g\(i-\frac{1}{2}\)\)^3+g^3\(i-\frac{1}{2}\)^3\]-\frac{g^2}{4}\sum_i n_i=g^2\,e^+_{4,n'}(1/g)\;.
 \end{equation}
 As it can be seen from the first equation (\ref{E_trois_g_sym}), the eigenfunctions of $\CI_r^\pm$ can be related by the change of sign of the bosonic operators. Comparing to the case $g=1$, we notice that the effect of this transformation is to transpose the Young diagram $n$ which labels the eigenstate.
We conclude by continuity in $g$ that this property holds at any $g$ and we have $e^+_{3,n}(g)=e^-_{3,n'}(g)$.
 
 In the classical limit $g\to 0$, the Hamiltonians (\ref{H_trois_g}) and (\ref{H_trois_g}) are the conserved quantities of the Benjamin-Ono equation.  Let us set $v=\sqrt{g}\partial \phi$ and $\CI_r\to g\CI_r$; in the classical limit we obtain
 \begin{eqnarray}
\CI_2&\sim&\int dx\ \frac{1}{2} v^2\;,\\
\CI_3&\sim&\int dx \(\frac{1}{3}v^3+\frac{1}{2}v H(v_x)\)\;,\\ \nonumber
\CI_4&\sim&\int dx \(\frac{1}{4}v^4+\frac{1}{4}v_x^2+ \frac{3}{4}v^2H(v_x)\)\;,\\ \nonumber
 \end{eqnarray}
 where $v_x=\partial_x v$ and $H(f)$ is the Hilbert transform of the function $f$. $\CI_2\sim L_0$ corresponds to the stress-energy tensor. One can verify directly that the above quantities are integrals of motion of the Benjamin-Ono equation
\begin{equation}
\dot v=v v_x+ \frac{1}{2}H(v_{xx})\;.
 \end{equation}
The conservation of $\CI_4$ relies on the identity $\int dx f^3=3\int dx f H(f)^2$ applied to $f=v_x$. 
 
The full integral of motion $I_3^\pm(g)$ for the $u(1)\otimes Vir(g)$ component from (\ref{I3vir}) can be written as
\begin{eqnarray}
I^\pm_3(g)&=&\CI_3^\pm(c;g)+\CI_3^\pm(\tilde c;g)\pm\sqrt{2g}\,(b_0-\alpha_0)(\CI_2(c)-\CI_2(\tilde c))+\nonumber \\
&+& (1-g)\[(1\mp1)\sum_{m>0}m\tilde c_{-m}c_m+(1\pm1)\sum_{m>0}mc_{-m}\tilde c_m\]\;.
\label{I3triangMM}
 \end{eqnarray} 
 The first line of  this formula is an operator which can be diagonalized in the basis of Jack polynomials spanned by
     \begin{equation}
    \label{jbasis}
J^{1/g}_{n^o} (\tilde c)\; J^{1/g}_{n^e}(c)\,|q\rangle+J^{1/g}_{n^o} ( c)\; J^{1/g}_{n^e}(\tilde c)\,|2\alpha_0-q\rangle
 \end{equation}
 while the second line has a triangular structure in this basis, in the sense that it removes bosons of one type and it creates bosons of the other type. Here we have incorporated the reflexion property which is built in in (\ref{I3triangMM}) and which  in this case exchanges the two copies of bosons and simultaneously reflects the charge, $q \to 2\alpha_0-q$.
 Due to the triangularity property, we conclude that the energy can be written as a sum 
 \begin{eqnarray}
 \label{E_trois_sum}
E^\pm_{3;\,n^o,n^e}(g)=e^\pm_{3,n^o}(g)+e^\pm_{3,n^e}(g)\mp\sqrt{2g}\,(q-\alpha_0)(|n^o|-|n^e|) \;.
\end{eqnarray} 
For the module of the identity $q=0$, comparing with (\ref{en_deux_part_MM}) we find that
 \begin{eqnarray}
 \label{E_trois_comp}
E^+_{3;\,n^o,n^e}(g)=\CE^\alpha_\lambda-\CE^\alpha_{\Lambda^0}+\(\frac{N-2}{\alpha}\)\(\CP_\lambda-\CP_{\Lambda^0}\)  + (M-2)(|n^e|+|n^o|)\;,
\end{eqnarray} 
where we remind that $\alpha=1/(1-g)$.
The  terms depending on the total momentum  come from the redefinition of the Hamiltonian, see for example formula (\ref{const_MM}).
This relation  proves the ansatz used in  (\ref{two_young}) and  (\ref{two_young_dual}).

 \newsec{$\textrm{WA}_{k-1}$ theories}
 \label{WA_section}
 
The duality obeyed by the conformal blocks was first discovered \cite{EBS} in the context of $\textrm{WA}_{k-1}(k+1,k+r)$ theories, which for $r=2$ are related to $\mathbb{Z}_k$
parafermions. For $k=2$ these models coincide with the Virasoro models from the section \ref{sec:MM}, with $g=(2+r)/3$. When $r$ is integer, the
$\textrm{WA}_{k-1}(k+1,k+r)$ theories correspond to $\mathbb{Z}_k^{(r)}$ parafermions considered in \cite{ES,EBS,BernevigHaldane1,BernevigHaldane2,BernevigW,JM} . The results in \cite{EBS} generalize straightforwardly to any value of $r$, not necessarily integer, in the same manner the results for the Ising CFT were extended to  generic Virasoro models in the previous section. 
Again, the  object under consideration is the dressed conformal block  
\begin{equation}
 \mathcal{F}^{a,b}_{M,N}(w;z)\equiv\langle \sigma(w_1)\dots \sigma(w_M)\Psi(z_1)\dots \Psi(z_N)\rangle^{a,b} \prod_{1\leq i <j}^{M}w_{ij}^{\frac{\tilde r}{k}} \prod_{i,j}(w_i-z_j)^{\frac{1}{k}} \prod_{1\leq i<j}^{N}z_{ij}^{\frac{r}{k}}
 \label{qhWA}
 \end{equation}
 where now $\sigma(w)$ and $\Psi(z)$ represent the primary fields $\Phi_{1,\dots,1,2|1,\dots1}(w)$ and $\Phi_{1,\dots,1|2,1,\dots1}(z)$ 
 with conformal dimensions  $\tilde r (k-1)/2k$ and 
 $r(k-1)/2k$ respectively, where $\tilde r$ is implicitly defined in equation (\ref{ccwa}) below.

As it was shown in reference \cite{EBS}, the dressed conformal block defined in equation  (\ref{qhWA})  obeys the second order differential equation
\begin{eqnarray}
\alpha\; h^{\alpha}(z)\;\mathcal{F}^{a,b}_{M,N}(w;z)=\tilde \alpha\;h^{\tilde \alpha}(w)\;\mathcal{F}^{a,b}_{M,N}(w;z)
\label{dec_eq_WA}
\end{eqnarray}
where $ h^{\alpha}(z)$ and $h^{\tilde \alpha}(w)$ are defined in terms of two differential Calogero-Sutherland operators 
\begin{eqnarray}
 h^{\alpha}(z)= \mathcal{H}^{\alpha}(z)-\E^{\alpha}_{0}+\(\frac{N-k}{ \alpha}-1\) [\mathcal{P}(z)-\mathcal{P}_0]-\frac{NM(M-k)}{k^2}\;,\\
  h^{\tilde \alpha}(w)= \mathcal{H}^{\tilde\alpha}(w)-\E^{\tilde\alpha}_0+\(\frac{M-k}{ \tilde\alpha}-1\) [\mathcal{P}(w)-\mathcal{P}'_0]-\frac{NM(N-k)}{k^2}\;.
\label{dec_eq_PFbis}
\end{eqnarray}
The coupling constant take now the values
\begin{equation}
\label{ccwa}
\alpha=-\frac{k+1}{r-1}\;, \qquad \tilde \alpha =\frac{k+r}{r-1}\;, \quad {\rm and} \quad g\equiv -\frac{\tilde \alpha}{\alpha}=\frac{k+r}{k+1} =\frac{k+1}{k+\tilde r}\;.
\end{equation}
The constants $\E^{\alpha}_0$ and $\mathcal{P}_0$ are given by
\begin{equation}
\E^{\alpha}_0\equiv \E^{\alpha}_{\lambda^0}=\frac{rN(N-k)[2Nr+k^2(1-2r)+k(N-r+Nr)]}{6k^2(k+1)}\;, \qquad \mathcal{P}_0=\frac{rN(N-k)}{2k}\;,
\end{equation}
while $\E^{\tilde \alpha}_0$ and $\mathcal{P}'_0$ are given by similar expressions with $r \to \tilde r$, which is equivalent to $g\to g^{-1}$, and $N\to M$.
The degree of homogeneity of  $\mathcal{F}^{a,b}_{M,N}(w;z)$ in both the variables $w$ and $z$ is 
\begin{equation}
\mathcal{P}(z)+\mathcal{P}(w)=\frac{rN(N-k)}{2k} + \frac{\tilde r M(M-k)}{2k}+\frac{MN}{k} \;.
\end{equation}
In the following we suppose that both $M$ and $N$ are divisible by $k$, which insures that the conformal block  (\ref{qhWA}) is non-zero. The duality property
(\ref{dec_eq_WA}) implies that the conformal block  (\ref{qhWA}) can be expanded on eigenfunctions of the dual Calogero-Sutherland Hamiltonians similarly to (\ref{dual_exp_MM}), 
\begin{equation}
\label{dual_exp_WA}
\mathcal{F}^{a,b}_{M,N}(w;z)=\sum_\lambda \CF_{\lambda'}^{\tilde \alpha,a}(w)\,\CF_{\lambda}^{\alpha,b}(z)\;.
\end{equation}
The lowest eigenstate of the Hamiltonian  $h^{\tilde \alpha}(w)$ is characterized now by the quantum numbers
     \begin{eqnarray}
 \lambda'^{0}_{kj-k+1}= \ldots=\lambda'^{0}_{kj}=\tilde r\(\frac{M}{k}-j\)\;, \quad j=1,\dots,M/k\;.
  \end{eqnarray}
  and it corresponds under the duality to the maximum ``partition'' $\Lambda^0$ defined as
    \begin{eqnarray}
      \label{def_lambda_max_WA}
  \Lambda^0_i&=&\lambda^0_i+\frac{M}{k}\;, \quad i=1,\dots,N\\   \nonumber
   \lambda^{0}_{ki-k+1}=\ldots=\lambda^{0}_{ki}&=&r\(\frac{N}{k}-i\)\;, \quad i=1,\dots,N/k\;.
  \end{eqnarray}
Generically, the sets $\lambda$ and $\lambda'$ are related to each other through a set of $k$ partitions $n^{(p)}$ and their duals $n'^{(p)}$, with $p=1,\dots,k$,
  \    \begin{eqnarray}
  \label{id_part_W}
\lambda_{ki-k+p}=\Lambda^{0}_{ki-k+p}-n^{(k-p+1)}_{N/k-i+1}\;, \qquad 
\lambda'_{kj-k+p}=\lambda'^{0}_{kj-k+p}+n'^{(p)}_{j}\;, \quad \quad p=1,\dots,k\;.
  \end{eqnarray}
Expressed in terms of the partitions  $n^{(p)}$ and $n'^{(p)}$, the energies of the intermediate state in the expansion (\ref{dual_exp_WA})
are
  \begin{eqnarray}
\E_\lambda^{\alpha}-\E_{\Lambda^0}^{\alpha}=
\label{en_deux_part_WA}
\sum_{p=1}^k\[b(n'^{(p)})-{g}\,b(n^{(p)})\]+\sum_{p=1}^k\[ 
(1-g)(N-2(p-1))-\frac{2M}{k}+g
\]|n^{(p)}|\;
\label{eig_l_WA}
 \end{eqnarray} 
and
  \begin{eqnarray}
\E_{\lambda'}^{\tilde \alpha}-\E_0^{\tilde \alpha}
=\sum_{p=1}^k\[b(n^{(p)})-\frac{1}{g}\, b(n'^{(p)})\]+\sum_{p=1}^k\[ 
\frac{2M\tilde r}{k}+(2-3g^{-1})+(1-g^{-1})(M+2(k-p))
\]|n'^{(p)}|\;, 
\label{eig_lp_WA}
 \end{eqnarray}   
where $b(n)$ and $|n|$ are defined in (\ref{bstan}). The two dual energies sum up to a constant which does not depend on the particular state $n^{(p)}$, as implied
by the formula (\ref{dec_eq_WA}),
 \begin{eqnarray} 
&\ &\E_\lambda^{\alpha}-\E_0^{\alpha}+g\,(\E_{\lambda'}^{\tilde \alpha}-\E_{0}^{\tilde \alpha})+\(\frac{N-k}{\alpha}-1\)\({\cal P}_\lambda-{\cal P}_0\)+g\(\frac{M-k}{\tilde \alpha}-1\)\({\cal P}_{\lambda'}-{\cal P}_{0'}\)= \nonumber \\ 
&\ & =\E_{\Lambda^0}^{\alpha}-\E_0^{\alpha}+\frac{NM}{k}\(\frac{N-k}{ \alpha}-1\)=\frac{NM(M-k)}{k^2}+\frac{gMN(N-2)}{k^2}\;.
 \end{eqnarray} 
 Therefore, the only difference with the Virasoro model from section \ref{sec:MM} is the appearance in the intermediate states of $k$ partitions. In the folowing, we will explain this structure via the bosonisation.
  
 \subsection{Bosonisation of the  $\textrm{WA}_{k-1}$ theories}
 
  The $\textrm{WA}_{k-1}$ theories can be constructed with the help of a $k-1$ component bosonic field. 
 Let $h_i$ with $i=1,\ldots, k$ be the weights in the fundamental representation of the $su(k)$ algebra, 
    \begin{equation}
 \vec  h_1=\(\frac{k-1}{k},-\frac{1}{k}, \ldots -\frac{1}{k}\)\;, \quad \ldots\;,\quad  \vec h_k=\(-\frac{1}{k}, \ldots -\frac{1}{k}, \frac{k-1}{k}\)\;.
  \end{equation}
  The fundamental weights of $su(k)$ are 
      \begin{equation}
 \vec \omega_i=\sum_{j=1}^i \vec h_j\; \quad {\rm and} \quad \rho\equiv \sum_{j=1}^k \vec \omega_j=\(\frac{k-1}{2},\frac{k-3}{2}, \ldots,-\frac{k-1}{2}\)\;.
  \end{equation}
Let us first consider $k$ copies of bosonic fields  $\vec \phi =(\phi^1,\ldots, \phi^{k})$ with normalization
 $\langle\phi^i(z)\phi^j(0)\rangle=-\delta_{ij}\log z$. We redefine the fields as
  \begin{eqnarray}
 &\ &  \phi_0=\frac{1}{k}\sum_{i=1}^k \phi^i\;,\\
  &\ & \phi_i=\vec h_i \, \vec \phi=\phi^i-\phi_0 \;, \quad i=1,\ldots, k \nonumber
  \end{eqnarray}
    The $k$ fields $\phi_i$ are not independent since $\sum_{i=1}^k \phi_i=0$. They are the fields that effectively enter the bosonisation
    of the $\textrm{WA}_{k-1}$ theory. The diagonal field $\phi_0$ decouples at this stage, but it is convenient to keep it for later purpose.
    It will appear in the next subsection in guise of the $u(1)$ field.
    We have 
    \begin{eqnarray}
     &\ &  \langle\phi_0(z)\phi_0(0)\rangle=-\frac{1}{k}\log z\;, \\ \nonumber
 &\ &  \langle\phi_i(z)\phi_0(0)\rangle=0\;, \qquad i=1,\ldots, k\\ \nonumber
 &\ & \langle\phi_i(z)\phi_j(0)\rangle=-\(\delta_{ij}-\frac{1}{k}\)\log z\;.
  \end{eqnarray}
   The fields $\phi_i$ generate a conformal field energy with stress-energy tensor 
  \begin{equation}
T(z)=\sum_{i<j}: \partial \phi_i\partial \phi_j:-i{\alpha_0}{\sqrt{2}}\sum_j(j-1)\partial^2\phi_j=-\frac{1}{2}\sum_{j=1}^k:(\partial \phi^j)^2:+\frac{k}{2}\:(\partial \phi_0)^2:+i{\alpha_0}{\sqrt{2}}\vec \rho\,\partial^2\vec \phi\;
 \end{equation}
and a spin 3 current $\tilde W(z)$  given by\footnote{The usual normalization is such that $\langle W(1)W(0)\rangle=c/3$, so that $\tilde W(z)=
\sqrt{\frac{k-2}{2kg} ((k+2)g-k)((k+2)-kg)} W(z)$. }
 \begin{eqnarray}
i\tilde W(z)&=&\sum_{i<j<l}:\partial\phi_i\partial\phi_j\partial\phi_l: -{\alpha_0}{\sqrt{2}i}\sum_{j<l}\[(l-1):\partial^2\phi_j\partial\phi_l: +(j-2):\partial\phi_j\partial^2\phi_l: \]+\\ \nonumber
&-&{\alpha_0^2}\sum_j(j-1)(j-2)\partial^3\phi_j+\frac{i\alpha_0(k-2)}{\sqrt{2}}\partial T(z)\;.
 \end{eqnarray}
The vertex operators 
 \begin{equation}
 V_{\vec \beta}(z)=:e^{i\vec \beta\, \vec \phi (z)}:
 \end{equation}
 are primary fields of the theory, with conformal dimension given by\footnote{
 If the vector $\vec \beta$ is not orthogonal to the vector $(1,...,1)$, the vertex operator will have an extra $u(1)$ part, since it will contain $\phi_0$.}
 \begin{equation}
 \Delta_{\vec \beta}=\frac{1}{2}{\vec \beta(\vec \beta-2 \vec \alpha_0)}\;, \qquad \vec \alpha_0={\alpha_0}{\sqrt{2}}\ \vec \rho\;.
 \end{equation}
 The degenerate fields of the theory, denoted by $\Phi_{(n_1,...n_{k-1}|n'_1,...n'_{k-1})}$, are associated to the vertex operators with charges
 \begin{equation}
 \vec \beta=\frac{1}{\sqrt{2}}\sum_{i=1}^{k-1}[(1-n_i)\alpha_++(1-n'_i)\alpha_-]\,\vec\omega_i
 \end{equation}
 where
  \begin{equation}
 \alpha_{+}=\sqrt{\frac{2}{g}}\;, \quad \alpha_{-}=-\sqrt{{2}{g}}\;, \quad 2\alpha_0=\alpha_{+}+\alpha_{-}\;.
 \end{equation}
 so that the fundamental fields $\Phi_{(1,...,1,2|1,...,1)}$ and $\Phi_{(1,...,1|2,1,...,1)}$ are represented by the vertex operators 
$V_{-1/\sqrt{g} \vec \omega_{k-1}}(z)$ and $V_{\sqrt{g}\vec  \omega_1}(z)$ respectively,
 with conformal dimension
  \begin{equation}
\tilde h =\frac{k-1}{2k}\[\frac{k+1}{g}-k\]=\frac{\tilde r(k-1)}{2k}\;, \qquad h =\frac{k-1}{2k}\[(k+1)g-k\]=\frac{ r(k-1)}{2k}\;.
 \end{equation}

  The fields $\Phi_{(1,...,1|2,1,...,1)}$ have generically two fusion channels with themselves,
 \begin{equation}
\Phi_{(1,...,1|2,1,...,1)}(z_1)\Phi_{(1,...,1|2,1,...,1)}(z_2)\sim \frac{\Phi_{(1,...,1|1,2,...,1)}(z_2)}{(z_1-z_2)^{2h-h_a}}+
\frac{\Phi_{(1,...,1|3,1,...,1)}(z_2)}{(z_1-z_2)^{2h-h_s}}\;,
 \end{equation}
 so that the leading short distance singularity is characterized by the power $2h-h_a=r/k$. This explains the powers which were used in dressing the conformal block  (\ref{qhWA}) to remove the short distance singularities.

 Let us now consider the  $u(1)$ current $J(z)$ which we identify with the diagonal field $J(z)=i\sqrt{k}\partial \phi_0(z)$, We dress the fundamental fields
 $\Phi_{1,\dots,1|2,1,\dots1}(z)$ and $\Phi_{1,\dots,1,2|1,\dots1}(w)$   by vertex operators, defining
   \begin{eqnarray} 
  \label{vaPhiWA}
V (z)\equiv \Phi_{1,\dots,1|2,1,\dots1}(z) \,:e^{i\sqrt{g}\phi_0(z)} : \;, \qquad   \tilde V (w)\equiv \Phi_{1,\dots,1,2|1,\dots1}(w) \,:e^{i\frac{1}{\sqrt{g}}\phi_0(w)} :\;.
 \end{eqnarray}
Choosing appropriately the bosonic representative for the fundamental fields, we can write 
   \begin{eqnarray} 
  \label{vaPhiWAcomm}
V (z)\sim :e^{i\sqrt{g}\phi_1(z)}e^{i\sqrt{g}\phi_0(z)} :\, =\, :e^{i\sqrt{g}\phi^1(z)}:\;, \qquad   \tilde V (w)\sim :e^{i\frac{1}{\sqrt{g}}\phi_k(w)}e^{i\frac{1}{\sqrt{g}}\phi_0(w)} : \, =\, :e^{i\frac{1}{\sqrt{g}}\phi^k(w)}:\;,
 \end{eqnarray}
and which again can justify the mutual separation of the action of the Calogero-Sutherland Hamiltonians on the $u(1)$ dressed conformal blocks
(\ref{qhWA}). A similar basis of bosonic fields was used recently in \cite{KMS} to study the appearance of the $W_{1+\infty}$ algebra in the context
of the $su(3)$ AGT relationship.

\subsection{The $\textrm{WA}_{k-1}$ theories at $g=1$}
   
The action of the Calogero-Sutherland Hamiltonians on the conformal blocks can be again transferred to an action on the Hilbert space of the $u(1)\otimes \textrm{WA}_{k-1}$ as it does
in the case of minimal models, {\it cf.} section (\ref{corr}).
When $g=1$ the first non-trivial conserved quantity $I_3$ has the expression  
\begin{eqnarray}
I_3^{\pm}(1) &= &  \pm \frac{1}{\sqrt{k}}\[2 \sum_{m\neq 0}  a_{-m} L_{m} +   \sum_{m,k \geq 1} \left( a_{-m-k}a_m a_k  +  a_{-m}a_{-k}a_{m+k} \right) \] 
\pm \tilde {W_0}
 \end{eqnarray}
and  can be written as a sum of $k$ independent bosonic Calogero-Sutherland Hamiltonians.
The new ingredient $\tilde W_0$ is the zero mode of the operator $\tilde W(z)$ 
 \begin{equation}
i\tilde W(z)=\frac{1}{3} \sum_{j=1}^k :(\partial\phi_j)^3:=\sum_{j=1}^k \[\frac{1}{3} :(\partial\phi^j)^3:-
:(\partial\phi^j)^2(\partial\phi_0):\]+\frac{2k}{3}:(\partial\phi_0)^3:\;,
 \end{equation}
 while the Virasoro generators $L_m$ are the Fourier modes of the stress-energy tensor
 \begin{equation}
T(z)=-\frac{1}{2}\sum_{j=1}^k:(\partial \phi^j)^2:+\frac{k}{2}:(\partial \phi_0)^2:\;.
 \end{equation}

 Let us now identify the  $u(1)$ current with the diagonal bosonic field $J=i\sqrt{k}\partial \phi_0(z)$, with Fourier modes $a_m=\frac{1}{\sqrt{k}}\sum_{j=1}^kc_m^j$. It is now straightforward
 to show that $I_3$ can be written as a sum of $k$ decoupled Hamiltonians depending on the $k$ independent bosons $c_m^j$
 \begin{eqnarray}
I_3^{\pm}(1)=\pm\sum_{j=1}^k\CI_3(c^j)\pm2\sum_{j=1}^k (\vec h_j \vec c_0)\;\CI_2(c^j)\;,
 \end{eqnarray}
 with the only coupling between the $k$ bosonic copies being realized by the zero modes $c_0^j$ in the second term.
This decomposition generalizes the result of Belavin and Belavin \cite{BB} to the case of $W$ algebras with $g=1$, and it justifies the
structure of the eigenenergies of the intermediate states 
(\ref{en_deux_part_WA}) as a sum over $k$ Young tableaux.
 
\subsection{Arbitrary $g$}
 
When $g\neq1$, the integral of motion $I_3$ of the $W$ theories conserves the same triangular structure as the one described in section
\ref{minmodggen}. Its expression is given by 
  \begin{eqnarray}
  \nonumber
I_3^{\pm}(g) = k(1-g)\sum_{m\geq 1} m a_{-m}a_m  \pm 2\sqrt{\frac{g}{k}} \sum_{m\neq 0}  a_{-m} L_{m}  \pm  \sqrt{\frac{g}{k}} \sum_{m,k \geq 1}\left(  a_{-m-k}a_m a_k  +  a_{-m}a_{-k}a_{m+k} \right) 
\pm\sqrt{g} \tilde {W_0}\;.\\
\end{eqnarray} 
After expressing the zero mode of the $W(z)$ current in terms of the bosonic fields, one finds that
 \begin{eqnarray}
I_3^{\pm}(g)& =&\sqrt{g}\,I_3(1)+ k(1-g)\sum_{m\geq 1} m a_{-m}a_m\pm(1-g)\sum_{j<l}\sum_m m:c^j_{-m}c_m^l:+{\rm \ terms\ with\ zero\ modes}=\nonumber\\ 
&=&\sum_{j=1}^k \CI_3^\pm(c^j;g)\pm2\sqrt{g}\sum_{j=1}^k\, \vec h_j \cdot (\vec c_0-\vec \alpha_0)\; \CI_2(c^j)\mp \sqrt{g}\sum_j (\vec h_j\cdot \vec \alpha_0) (\vec h_j\cdot \vec c_0)(\vec h_j\cdot (\vec c_0-\vec \alpha_0))
\nonumber
\\ 
&+&(1-g)\sum_{m\geq 1} \[(1\pm1)\sum_{j<l}m:c^j_{-m}c_m^l:+(1\mp1)\sum_{j>l}m:c^j_{-m}c_m^l:\]\;,
\label{I3triangWA}
\end{eqnarray} 
where $\vec \alpha_0=\sqrt{2}\alpha_0\vec \rho$.
On states in the module of the identity, the eigenvalues of $I^+_3(g)$ are given by
 \begin{eqnarray}
E^+_{3;\,n^{(p)}}(g)=\sum_{p=1}^k e^+_{3,n^{(p)}}(g)+(1-g)\sum_{p=1}^k (k+1-2p) |n^{(p)}| \;,
\end{eqnarray} 
where we have ordered the partitions in the reverse order, in the sense that the partition $n^{(p)}$ corresponds to the boson copy 
$c^{k-p}$. This eigenenergy agrees, up to a momentum-depending shift in the Hamiltonian, with the one in (\ref{en_deux_part_WA}),
\begin{eqnarray}
 \label{E_trois_comp_W}
E^+_{3;\,n^{(p)}}(g)=\CE^\alpha_\lambda-\CE^\alpha_{\Lambda^0}+\(\frac{N-k}{\alpha}\)\(\CP_\lambda-\CP_{\Lambda^0}\)  + \frac{2(M-k)}{k}\sum_{p=1}^k |n^{(p)}|\;.
\end{eqnarray} 
This proves the identification of the labels of the "partitions" $\lambda$ in terms of $k$ partitions $n^{(p)}$ which was done in (\ref{id_part_W}).

\newsec{Conclusion and outlook}

In this paper we have studied the integrable structure of the $u(1)\otimes Vir(g)$ and, more general, of the $u(1)\otimes \textrm{WA}_{k-1}$ CFTs. Our starting point is the action of the Calogero-Sutherland Hamiltonian on the conformal blocs of these theories which contains second order degenerate fields. Dual degenerate field are associated to Calogero-Sutherland Hamiltonians with dual coupling constants. Once translated on arbitrary descendant fields inserted in the correlation functions, the action of the Calogero-Sutherland Hamiltonians generates an action on the Hilbert space of the theory. This action corresponds to the action of the integrals of motion found in refs. \cite{AFLT} and \cite{FL11}. The basis associated to these integrals of motion were used in refs. \cite{AFLT} and \cite{FL11} to give a proof of the AGT conjecture \cite{AGT} relating the Nekrasov's instanton partition function for supersymmetric quiver gauge theories to Liouville conformal blocks (or to the conformal blocks of the $\textrm{WA}_{k-1}$ theory). Using bosonisation of the corresponding theories, we show that the action of the Calogero-Sutherland integrals of motion on the Hilbert space can be written as a sum of $k$ copies of Calogero-Sutherland bosonic Hamiltonians coupled by an interaction term which is triangular in the bosonic basis. This explains why the spectrum of the integrals of motion is the sum of the spectra of $k$ Calogero-Sutherland Hamiltonians with coupling constant $g$ (or, dually, $1/g$) and it is indexed by $k$ partitions. 
The advantage of our approach is to show that the integrals of motion correspond to a unique Calogero-Sutherland differential operator with coupling constant $1-g$, and that the associated eigenfunctions are generically non-polynomial. In some particular cases, when $g=(k+r)/(k+1)$ with $k$ and $r$ integers, the associated eigenfunctions are Jack polynomial with parameter $\alpha=-(k+1)/(r-1)$. These polynomials give electron eigenfunctions of the Fractional Quantum Hall Effect (FQHE) with pairing properties \cite{ES, EBS}.

It would be interesting to explore further the non-polynomial eigenfunctions of the Calogero-Sutherland model which are associated with the conformal blocks, and their link with the 
bosonic states in their Jack polynomial representation.  

One of the most interesting yet unsolved problem is to unravel the integrable structure which is behind the integrals of motion which we identified in this work. We have obtained that the integrals of motion are associated with particular null vectors which are descendants of the null vector at level two. It would be interesting to find a representation of the 
monodromy matrix in the framework of the CFT, similar to that obtained by Bazhanov, Lukyanov and Zamolodchikov \cite{BLZ} for  $Vir(g)$. 
 
 \bigskip 

{\bf Note}: On the final stage of the preparation of the manuscript we have learned that results concerning the integrals of motion
$I_n$ and the associated $R$ matrix were obtained by Davesh Maulik and Andrei Okounkov \cite{DOk}. Our results partially overlap with with theirs,
in particular concerning the triangular expansion of $I_3$, our formulas (\ref{I3triangMM}) and  (\ref{I3triangWA}). 
An expression similar to (\ref{I3triangMM}) also appeared recently in \cite{SWY}. A $q$-version of the AGT relation was given in \cite{SHIRA}.
The proof of the AGT conjecture for $U(k)$ quiver gauge theories was given 
by Fateev and Litvinov in \cite{FL11}, where they used the same basis of the integrals of motion as the one in our section \ref{WA_section}. 
\\
  
{\bf Acknowledgements}: V.P and D.S. thank to Andrei Okounkov for discussions and for sharing their unpublished work. We acknowledge discussions with A.A.~Belavin, Vl.~Dotsenko,  B.~Feigin, A.A.~Litvinov,  K.~Schoutens, J.~Shiraishi, M.~Talon and P.~Wiegmann.  B.E. was supported by the foundation FOM of The Netherlands.

\appendix

\newsec{$u(1)\otimes \mbox{Ising}$  conformal field theory}
In section (\ref{sec:MM}) we considered  the $u(1)\otimes Vir(g)$ algebra and the corresponding family of  CS conformal block eigenfunctions. In particular we derived the CS eigenvalue formulas (\ref{dual_energy-MM}) which show how the duality $g\to 1/g$ of the CS model works for non-polynomial eigenfunctions. The formulas  (\ref{dual_energy-MM}) have been initially guessed on the basis of the results concerning the  $u(1)\otimes Vir(4/3)$ algebra, that is to say the $u(1)\otimes \mbox{Ising}$ algebra.  This case  is particularly useful because  the conformal blocks  of free fermion fields (which represent  one side of the duality) are given by certain  Jack polynomials with pairing  properties.  It is interesting to remark that these properties  have a direct relevance in the study of the FQHE.
In this appendix we show in full detail that the  formulas  (\ref{dual_energy-MM}) can be derived, for the Ising case,  by using, together with the (\ref{dec_eq_MM}), the pairing properties of the fermionic conformal blocks.

\subsection{Ising primary fields}
The Ising model is the unitary minimal model of the  Virasoro algebra (\ref{viras}) with central charge $c=1/2$. It presents a finite number of operators which 
close under operator algebra: besides the  identity $\mbox{I}$,  
there are only  the two Virasoro primary fields 
$\Phi_{(1|2)}$ and $\Phi_{(2|1)}$ with conformal dimension $\Delta_{(1|2)}=1/2$ and $\Delta_{(2|1)}=1/16$. In our notations, this fixes $g=4/3$.
 The fields $\Phi_{(1|2)}$ and $\Phi_{(2|1)}$ correspond to the free fermion field $\Psi$:
 \begin{equation}
\Psi \equiv \Phi_{(1|2)}, \nonumber
\end{equation}
 and to the spin operator $\sigma$:
 \begin{equation}
\sigma\equiv \Phi_{(2|1)}.
 \end{equation}
 Their fusion relations read 
\begin{eqnarray}
\Psi(z) \Psi(w)&=&\frac{1}{z-w} \label{pp_ope}\,\mbox{I}\;, \\
\sigma(z) \sigma(w) &=& \frac{C_{\sigma,\sigma}^{I}}{(z-w)^{1/8}}\; \mbox{I}+\frac{C_{\sigma,\sigma}^{\Psi}}{(z-w)^{-3/8}} \Psi(w) \label{ss_ope}\;, \\
\sigma(z) \Psi(w)&=&\frac{C_{\sigma,\psi}^{\sigma}}{(z-w)^{1/2}} \sigma(w)\label{sp_ope}\;,
\end{eqnarray}  
where the $C_{X,Y}^{Z}$ are the structure constants of the operator algebra.
\subsection{Relation with Calogero-Sutherland model}
\label{Ising_cs}

{\bf Conformal block of $N$ free fermions : }
Consider first the conformal block of $N$ free fermions:
\begin{equation}
  \langle  \Psi(z_1)\ldots \Psi(z_N)\rangle.
 \end{equation} 
This correlator can of course be easily computed   by using the Wick theorem,  $\langle  \Psi(z_1)\dots \Psi(z_N)\rangle=\mbox{Pf}(1/z_{ij})$, where $\mbox{Pf}(M_{ij})$ if the Pfaffian of the matrix $M_{ij}$.  The $N$ fermion fields  $\Psi$   degenerate second order fields and  their correlation function  satisfies  a system of $N$  second order equations (\ref{12_diff_eq}). On can recast these equations in the following way: 
\begin{equation}
\sum_{i=1}^{N} z_i^2\, \mathcal{O}^{4/3}_i  \langle  \Psi(z_1)\dots\Psi(z_N)\rangle=0\;.
\label{Nff_de}
\end{equation}
Using the Ward identities satisfied by the conformal blocks
\begin{eqnarray}
 \sum_{i=1}^{N}\partial_{i}  \, \langle  \Psi(z_1)\dots\Psi(z_N)\rangle&=&0\nonumber \\
\sum_{i=1}^{N} \left(z_i \partial_{i}+ \frac{1}{2}\right)  \langle  \Psi(z_1)\dots\Psi(z_N)\rangle&=&0,
 \end{eqnarray}
one obtains from the equation (\ref{Nff_de}):
\begin{equation}
\left[\sum_{i=1}^N(z_i \partial_i)^2-\frac{4}{3}\sum_{ i<j} \frac{z_i z_j}{z_{ij}^2}+\frac{2}{3}\sum_{ i<j}\frac{z_i+z_j}{z_{ij}}(z_i\partial_i-z_j\partial_j) -\frac{2N}{3}\right] \langle  \Psi(z_1)\dots\Psi(z_N)\rangle=0\;.
\label{cs_eq_pp}
\end{equation}
One can recognize in the above  the  Calogero-Sutherland Hamiltonian in the form (\ref{cs_gamma}) with  $\gamma=2/3$ and $g=4/3$. The function 
\begin{equation}
\Psi(z)\equiv \prod_{ i<j} z_{ij}^{2/3}\  \mbox{Pf}\left(1/z_{ij}\right)
\label{free_fer_pf}
\end{equation}
 is then eigenfunction of (\ref{CS_g_0}) with coupling $g=4/3$. It is easy to convince oneself that
 \begin{equation}
 \Psi(z)=\prod_{i<j} z_{ij}^{-1/3}\; F(z)=\prod_{i<j} z_{ij}^{1-g}\; F(z)\;,
 \end{equation}
 with $F(z)$ regular at $z_i\to z_j$, which means that $\Psi(z)$ is subject to the second type of boundary conditions $\Psi^-(z)$ in the terminology of equation (\ref{eig_bc}).
 \bigskip
 
\noindent {\bf Conformal blocks of $M$ spin operators $\sigma$ :} Analogously, we could consider the correlation function of $M$ fields $\sigma$:
\begin{equation}
  \langle  \sigma(z_1)\dots \sigma(z_M)\rangle_a
  \label{ss_gs}
\end{equation}
where $a=1,\dots ,2^{M/2-1}$ is the conformal block index. Indeed,  from the fusions (\ref{ss_ope})-(\ref{sp_ope}), there are  $2^{M/2-1}$ different conformal blocks corresponding to the function (\ref{ss_gs}).

\begin{figure}[h]
\begin{center}
      \includegraphics[scale=0.4,angle=-0]{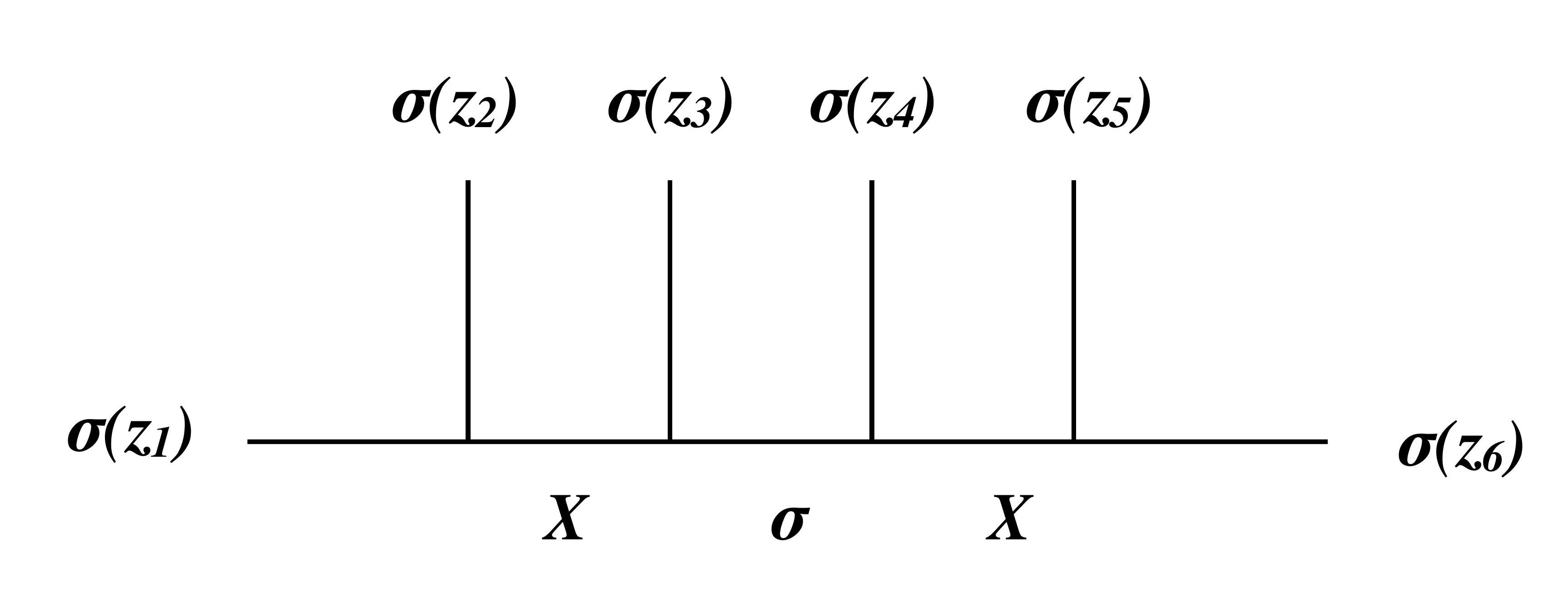}
      \end{center}
    \caption{ A diagram representing the  conformal block of  $\sigma$ fields for $M=6$.  For each diagram there are $M/2-1$ fields $X$ 
which can correspond to the $I$ or to the $\psi$ field, 
with $X=\mbox{I}$ or $X=\psi$. The total number of possible conformal blocks is then $2^{M/2-1}$.} 
   \label{cblock}  
   \end{figure}

Again, using the $M$  second order differential equations  (\ref{12_diff_eq}) and  the conformal  Ward identities, the conformal block (\ref{ss_gs}) can be shown to satisfy  the equation:
\begin{equation}
\left[\sum_{i=1}^M(z_i \partial_i)^2-\frac{3}{32}\sum_{ i< j} \frac{z_i z_j}{z_{ij}^2}+\frac{3}{8}\sum_{ i<j}\frac{z_i+z_j}{z_{ij}}(z_i\partial_i-z_j\partial_j) -\frac{3 M}{64} \right] \langle  \sigma(z_1)\dots \sigma(z_N)\rangle_{a}=0,
\label{cs_eq_ss}
\end{equation}
which corresponds to the operator (\ref{cs_gamma}) for $\gamma=3/8$ and $g=3/4$.  An  eigenfunction of (\ref{CS_g_0}) with coupling $g=3/4$ is then obtained by setting:
\begin{equation}
\Psi(z)_{a}\equiv \prod_{ i<j} z_{ij}^{3/8}  \langle  \sigma(z_1)\dots \sigma(z_M)\rangle_a.
\label{es_cs_qh}
\end{equation} 
It is interesting to notice that the eigenvalue associated to the eigenfunctions $\Psi(z)_{a}$ does not depend on the particular conformal
block index $a$. 
Generally, by using the (\ref{ss_ope}) into (\ref{es_cs_qh}), one has that:
\begin{equation}
\Psi(z)_{a}\sim c_{a1} \,z_{i j}^{1/4}+c_{a2} \,z_{i j}^{3/4}\;\; \quad \mbox{for} \quad z_i\to z_j.
\end{equation}
 The possible asymptotic behavior characterizing  the  eigenfunctions of (\ref{CS_g_0}) with $g=3/4$,  see (\ref{bcs}),  are then associated  to the two fusion channels in (\ref{ss_ope}). The exponents characterizing the two boundary condition are given by $1-g=1/4$ and $g=3/4$. The first exponent is smaller, so we can write
 \begin{eqnarray}
\Psi(z)_{a}=\prod_{i<j} \,z_{i j}^{1/4}\;F(z)_a\;\; \\ \nonumber
 {\rm with}\quad F(z)_a\sim c_{a1}+c_{a2}\sqrt{z_{ij}} \quad \mbox{for} \quad z_i\to z_j.
\end{eqnarray}
Each conformal block $\Psi(z)_{a}$ is characterized  by having a given configuration of boundary conditions. In this respect,  consider for instance  the simplest non trivial case, i.e. with $N=4$. Here one has  two conformal blocks, $\Psi(z)_{a}$ with $a=1,2$. One conformal block, say  $\Psi(z)_{1}$, can be chosen such that:
\begin{align}
&\Psi(z)_1\sim_{z_1\to z_2} z_{12}^{1/4} & & \Psi(z)_1\sim_{z_3\to z_4} z_{34}^{1/4}  \nonumber \\
& \Psi(z)_1 \sim_{z_1\to z_3} c_{11} \,z_{13}^{1/4}+c_{12} \,z_{13}^{3/4} & &  \Psi(z)_1 \sim_{z_2\to z_4} c_{11}\, z_{24}^{1/4}+c_{12} z_{24}^{3/4} \nonumber
\end{align}
while the other behave as:
\begin{align}
&\Psi(z)_2  \sim_{z_1\to z_2}  z_{12}^{3/4} & &  \Psi(z)_2  \sim_{z_3\to z_4} z_{34}^{3/4}  \nonumber \\
& \Psi(z)_2  \sim_{z_1\to z_3}  c_{21}\, z_{13}^{1/4}+c_{22}\, z_{13}^{3/4}      & &  \Psi(z)_2   \sim_{z_2\to z_4} c_{21}\, z_{24}^{1/4}+c_{22} \,z_{24}^{3/4} \nonumber
\end{align}
where the  $c_{nm}$ are some constants. A detailed discussion about the possible boundary conditions configurations associated to conformal block correlator has been done in \cite{Cardy_Doyon}.

\subsection{Clustering polynomials and admissible partitions}
 Let us come back for a moment to the fermionic conformal blocks.
 The function 
\begin{equation}
\prod_{i<j} z_{ij}\; \langle\Psi(z_1)\dots \Psi(z_N)\rangle 
\label{MR_gs}
\end{equation}
is a function regular when $z_{ij}\to 0$, monovalued, symmetric of total degree $N(N-2)/2$, therefore it should be a symmetric polynomial\footnote{$N$ should be even, otherwise the conformal block vanishes.}.
It is an eigenfunction of the CS Hamiltonian (\ref{CS_g_g}) with $\alpha=1/(1-g)=-3$, therefore it should be a Jack polynomial with a negative coupling constant. By inspection it is equal to $J^{-3}_{\lambda_{0}}(z)$
where $\lambda_0$ is the partition
\begin{equation}
\lambda_{0}=[N-2,N-2, N-4,N-4,\cdots ,0, 0]\;.
\label{pf_l0}
\end{equation}
It is interesting to point out that the polynomial $J^{-3}_{\lambda_{0}}(z)$  does not vanish when two variables are at  the same point  but vanishes with power $2$ when  the third  particles  approches a cluster of two.   Due to this property, this polynomial  is the zero-energy eigenstate of model $3$-body Hamiltonian and thus  it has been considered as a good  trial many-body wavefunctions  for fractional quantum Hall systems.  

More generally,  Jack polynomials with $(k,r)$-clustering properties  appear in the $\textrm{WA}_{k-1}(k+1,k+r)$ theories; these polynomials  vanish 
with a power $r$ when at least $k+1$ particles come to the same point.  
A  characterization of symmetric polynomials with  clustering properties was initiated in the work of Feigin {\it et al.} \cite{FJMM,FJMM2}. 
Let $k,r$ be positive integer such that $k+1$ and $r-1$ are co-prime. A  partition $\lambda$ is said to be $(k,r,N)$-admissible if it satisfies the following condition:
\begin{equation}
\lambda_{i}-\lambda_{i+k}\geq r \quad (1\leq i\leq N-k).
\label{admissible_part}
\end{equation}
The (\ref{pf_l0}) is then a $(2,2,N)$ admissible partition.
Given a $(k,r,N)$-admissible partition $\lambda$ Feigin {\it et al.} \cite{FJMM} showed that:

\begin{itemize} 
\item the coefficients $c_{\lambda \mu}(\alpha)$ do not have a pole for the particular negative value $\alpha=-(k+1)/(r-1)$.

\item the  Jack  polynomial $J^{-(k+1)/(r-1)}_{\lambda}(z_1,\cdots,z_n)$ vanishes when $z_1=z_2=\cdots=z_{k+1}$. 

\end{itemize}
The space spanned by the Jack polynomials $J^{-(k+1)/(r-1)}_{\lambda}(z_1,\cdots,z_N)$ for all $(k,r,N)$-admissible partitions $\lambda$ coincides with the space of symmetric polynomials satisfying the $(k,r)$ clusterings. 

\subsection{Duality and separation}
\label{sec:dual_sep_Ising}
Let us now consider  the function (\ref{qhMM}) for the Ising case:
 \begin{equation}
 \mathcal{F}^{a}_{M,N}(w;z)\equiv\langle \sigma(w_1) \cdots \sigma(w_M)\psi(z_1) \cdots \psi(z_N)\rangle_a \prod_{1\leq i <j}^{M}w_{ij}^{1/8} \prod_{i,j}(w_i-z_j)^{1/2} \prod_{1\leq i<j}^{N}z_{ij}.
 \label{qhIsing}
 \end{equation}
Note that,  with respect to the (\ref{qhMM}),  we have here  only  one index $a$ which runs over the possible $2^{M/2-1}$ possible independent conformal blocks (see also the (\ref{qhMM}) below). This is because the $\Psi$ field correlators have an Abelian monodromy.  The function 
$\mathcal{F}^{a}_{M,N}(w;z)$,  which has been introduced to describe the  
the excited $M-$quasihole   wavefunction for  the paired fractional quantum Hall state \cite{MooreRead,Nayak_Wilczek},   has  been computed exactly in \cite{Gurarie_is,Ardonne_is} (note that it  vanishes for $M$ odd).

 Because of the  fusions (\ref{pp_ope})-(\ref{sp_ope}), the factors  $\prod_{i<j} z_{ij}$  and $\prod_{i,j} (w_i-z_j)^{1/2}$ insure  the  function 
 $\mathcal{F}^{a}_{M,N}(w;z)$  to be  a symmetric polynomial  in the $z$ variables. In particular the factor $\prod_{i,j} (w_i-z_j)^{1/2}$ renders the variables $z$ and $w$ mutually local. The factor $\prod_{i<j} w_{ij}^{1/8}$ supresses the divergence as $w_{ij}\to 0$. It is rather easy to show  from the fusion (\ref{pp_ope}) that the function $\mathcal{F}^{a}_{M,N}(w;z)$  satisfies the following $(2,2)-$clustering properties:
 \begin{equation}
  \mathcal{F}^{a}_{M,N}(w,z_1=z_{2}=Z,z_{3},z_{4},\cdots, z_{N})=\prod_{i=1}^{M}(w_i-Z)\prod_{i=3}^{N}(Z-z_i)^2 \mathcal{F}^{a}_{M,N-2}(w,z_{3},z_{4},\cdots, z_{N}).
 \label{z2_clus}
 \end{equation} 
 The function $\mathcal{F}^{a}_{M,N}(w;z)$ can in general  
 be expanded in symmetric polynomials of $z$,   each of which  satisfies the $(2,2)$-clustering condition  and  has  total degree $D$ such that:
 \begin{equation}
 \frac{N(N-2)}{2}\leq D \leq \frac{N(N-2)}{2}+\frac{N M}{2}. 
  \end{equation}
    This can be seen, for instance, 
from  the conformal block $\mathcal{F}^{1}_{M,N}(w;z)$  corresponding to the case where all the $\sigma_{i}$ fuse into the identity. 
In the  limit $w_{2n }\to w_{2n-1}\equiv W_{n}$, $n=1,\dots ,M/2$,  one has from (\ref{ss_ope}):
\begin{equation}
\mathcal{F}^{1}_{M,N}(w_1,\dots ,w_{M},z)\to\prod_{i=1}^{M/2}\prod_{j=1}^{N} (z_j-W_i) J^{-3}_{\lambda_0}(z)\quad \mbox{for}  \quad w_{2n },w_{2n-1}\to W_{n} \quad n=1,\dots ,M/2.
\end{equation}
Similar considerations  can be made for all the conformal blocks.  
  
  It is therefore natural to expand the function $\mathcal{F}^{a}_{M,N}(w;z)$  on the basis of  Jack polynomial $J_{\lambda}^{-3}(z)$ where $\lambda$ is a $(2,2,N)$  admissible partition:
\begin{eqnarray}
&\ &\mathcal{F}^{a}(w;z) = \sum_{\lambda} \mathcal{F}^{4,a}_{\lambda'}(w) \, J^{-3}_{\lambda}(z)  \\
&\ &\lambda \;\; (2,2,N)-\mbox{admissible}\;,\qquad \lambda_i^0\leq \lambda_i\leq \Lambda^0_i(M)\;,
\label{decomp_qh_isi}
\end{eqnarray}  
where $\lambda^{0}_{i}$ and the  maximum  admissible partition $\Lambda^0(M)$ are given  respectively  by  (\ref{def_l_o}), where one has to take $h=1/2$,  and by (\ref{def_lambda_max_g}).
This shows that, for the Ising case, the choice (\ref{def_l_o}) is a direct consequence of the (2,2)-clustering properties. 
In the above expression we have:
\begin{equation}
\mathcal{H}^{-3}(z)\;J^{-3}_{\lambda}(z) =\E^{-3}_{\lambda}\,J^{-3}_{\lambda}(z)
\label{dev_is}
\end{equation}
with the energies  $\E_{\lambda}^{-3}$  given  by  (\ref{dec_eq_MM}) with $\alpha=-3$. 

The  $\mathcal{F}^{4,a}_{\lambda'}(w)$ are  non-polynomial functions of the variables $w$.
Specifying the (\ref{dec_eq_MM}) for the Ising case, one finds that the $P_{\lambda'}^{4}(w)$ are  eigenstates of $\mathcal{H}^{4}$:
\begin{equation}
\mathcal{H}^{4}(w)\,\mathcal{F}_{\lambda'}^{4,a}(w)=\E^{4}_{\lambda'}\,\mathcal{F}_{\lambda'}^{4,a}(w)
\end{equation}
with
\begin{equation}
\frac{3}{4}\E^{-3}_{\lambda} +\E^{4}_{\lambda'}+\frac{5 - M - N}{4} |\lambda|=E(N,M).
\label{relat_dual}
\end{equation}
 Note that we have associated the eigenfunctions  $P_{\lambda'}^{4,a}(w)$ to an admissible partition $\lambda'$. Until  now this is a simply consequence of the expansion 
 \ref{decomp_qh_isi}. Using   the expression for $\E^{-3}_{\lambda}$ together with the structure of the $(2,2,N)$-admissible partition $\lambda$, one is lead to the formulas  ({\ref{dual_energy-MM}) and $(\ref{two_young_dual})$ with $\tilde{\alpha}=4$ and $\tilde{h}=1/16$.

The construction shown above  generalizes to the case of  the   $u(1)\otimes WA_{k-1}$ algebras, discussed in section \ref{WA_section}. Indeed, analogously to the case of Ising, one can prove  that, in the  parafermionic models $\textrm{WA}_{k-1}(k+1,k+r)$ ($k$ and $r$  integers and $k,r\geq 2$), the conformal blocks of $\Phi_{(1,..,1|2,1,..,1)}$ and $\Phi_{(1,..,1|1,1,..,2)}$ fields \footnote{note that these fields correspond to the parafermionic currents $\Psi_{1}$ and $\Psi_{k-1}$ generating the $Z_k$ symmetry}  are expressed in terms of Jack polynomials with generalized  $(k,r)-$ clustering properties  \cite{EBS}.  Using the decoupling equations (\ref{dec_eq_WA}) together with the structure of the $(k,r,N)-$admissible Young tableau, one is naturally lead to the formulas (\ref{eig_l_WA})-(\ref{eig_lp_WA}).

 \underline{{\bf An explicit Ising example: $M=4$, and $N=2$}}
 \label{app: Ising_explicit}

For sake of clarity,  we give here a full  explicit example of the above results by considering the conformal block with $M=4$ spins operators and $N=2$ energy operators.
In  this case one has two independent conformal blocks, $\mathcal{F}^{a}(w;z)$, with $a=1,2$.

  The sum (\ref{decomp_qh_isi})  runs over the partitions $\lambda : \;\[\emptyset \],\[1\],\[2\],\[1,1\],\[2,1\],\[2,2\]$.  The corresponding Jack polynomials $J^{-3}_{\lambda}$ 
  are eigenfunction of $\mathcal{H}^{4}$ with energy:

\begin{center}
$
\setlength{\extrarowheight}{0.2cm}
\begin{array}{ccccccc}
Function&  J^{-3}_{\[\emptyset \]}& J^{-3}_{\[1\]}& J^{-3}_{\[2\]} & J^{-3}_{\[1,1\]} & J^{-3}_{\[2,1\]}&J^{-3}_{\[2,2\]} \\
Eigenval&  0& \frac{2}{3}& \frac{10}{3} & 2 & \frac{14}{3}&8\\
|n^o, n^e\rangle& |\bl2\br,\bl2\br\,\rangle&|\bl2\br,\bl1\br\,\rangle& |\bl2\br,\bl\emptyset\br\,\rangle &|\bl1\br,\bl1\br\,\rangle& |\bl1\br,\bl\emptyset\br\,\rangle&|\bl\emptyset\br,\bl\emptyset\br\,\rangle\\
\end{array}
$
\end{center}

while the  $\mathcal{F}^{4,a}_{\lambda'}(w)$,  are eigenfunction of  $\mathcal{H}^{4}$ with eigenvalues:

\begin{center}
$
\setlength{\extrarowheight}{0.2cm}
\begin{array}{ccccccc}
Function&  \mathcal{F}^{4,a}_{\[\frac{1}{4}+1,\frac{1}{4}+1,1,1\]}&\mathcal{F}^{4,a}_{\[\frac{1}{4}+1,\frac{1}{4}+1,1,0\]}& \mathcal{F}^{4,a}_{\[\frac{1}{4}+1,\frac{1}{4},1,0\]} &\mathcal{F}^{4,a}_{\[\frac{1}{4}+1,\frac{1}{4}+1,0,0\] }& \mathcal{F}^{4,a}_{\[\frac{1}{4}+1,\frac{1}{4},0,0\]}&\mathcal{F}^{4,a}_{\[\frac{1}{4},\frac{1}{4},0,0\]} \\
Eigenval&  \frac{43}{8}& \frac{41}{8}& \frac{27}{8} & \frac{35}{8} & \frac{21}{8}&\frac{3}{8}\\
|n'^o, n'^e\rangle& |\bl1,1\br,\bl1,1\br\,\rangle&|\bl1,1\br,\bl1\br\,\rangle& |\bl1,1\br,\bl\emptyset\br\,\rangle &|\bl1\br,\bl1\br\,\rangle& |\bl1\br,\bl\emptyset\br\,\rangle&|\bl\emptyset\br,\bl\emptyset\br\,\rangle\\
\end{array}
$
\end{center}

Note that the relation (\ref{relat_dual}) is satisfied.

We found the explicit form $\mathcal{F}^{4,a}_{\lambda'}$, $a=1,2$,  explicitly.
\begin{eqnarray*}
\mathcal{F}^{4,a}_{\[\frac{1}{4},\frac{1}{4},0,0\]}(w)&=&\sqrt{\sqrt{w_{13}\,w_{24}}-(-1)^a \sqrt{w_{23}\, w_{14}}} \label{sol11} \\
\mathcal{F}^{4,a}_{\[\frac{1}{4}+1,\frac{1}{4},0,0\]}(w)&=&m_{\[1\]}(w)\; \mathcal{F}^{4,a}_{\[\frac{1}{4},\frac{1}{4},0,0\]}  (w)\\
\mathcal{F}^{4,a}_{\[\frac{1}{4}+1,\frac{1}{4}+1,0,0\] }(w)&=&\left(-\frac{1}{4}m_{\[2\]} (w)+\sum_{i} w_i^3 \partial_{w_i}  \right)\; \mathcal{F}^{4,a}_{\[\frac{1}{4},\frac{1}{4},0,0\]} (w)\\
\mathcal{F}^{4,a}_{\[\frac{1}{4}+1,\frac{1}{4},1,0\]}(w)&=&\left(\frac{1}{4}m_{\[2\]}(w) +\frac{1}{2}m_{1,1}(w)-\sum_{i} w_i^3 \partial_{w_i}  \right) \; \mathcal{F}^{4,a}_{\[\frac{1}{4},\frac{1}{4},0,0\]} (w) \\
\mathcal{F}^{4,a}_{\[\frac{1}{4}+1,\frac{1}{4}+1,1,0\]}(w)&=&m_{\[2,1\]}(w)\; \mathcal{F}^{4,a}_{\[\frac{1}{4},\frac{1}{4},0,0\]} (w) \\
\mathcal{F}^{4,a}_{\[\frac{1}{4}+1,\frac{1}{4}+1,1,1\]}(w)&=&m_{\[1,1,1,1\]}(w)\; \mathcal{F}^{4,a}_{\[\frac{1}{4},\frac{1}{4},0,0\]}  (w)
\label{sol15}
\end{eqnarray*}
where the $m_{\lambda}(w)$ are the symmetric monomial associated to the partition $\lambda$:
\begin{equation}
m_{\lambda}(\{z_i \})=\mathcal{S} (\prod_i^N z_i^{\lambda_i}). 
\end{equation}
Here the $\mathcal{S}$ stands for the symmetrization over the $N$ variables.

  \newsec{Correspondence between CS Hamiltonians and Integral of motions}
 \label{Correspondence}
 
In this appendix we derive the correspondence of section \ref{corr}. This is the central result of this paper. The system of integrals of motions that we obtained is the same as the one introduced in \cite{AFLT}.

 \subsection{Algebraic setting}

We consider the algebra $Vir (g)\otimes \textrm{H}$ of central charge $c = 2 - 6(g-1)^2/g$ generated by
\begin{eqnarray}
&&[L_n,L_m] = (n-m)L_{n+m} + \frac{c}{12}n(n^2-1)\delta_{n+m,0} \\
&& [a_n,a_m] = n\delta_{n+m,0} \\
  &&[L_n,a_m]=0 
\end{eqnarray}
The Heisenberg algebra contains a Virasoro with $c=1$:
\begin{eqnarray}
& & l_n =  \frac{1}{2}\sum_{m \in \mathbb{Z}} a_{n-m}a_m \qquad n \neq0 \\
& & l_0 = \sum_{m >0} a_{-m} a_m + \frac{1}{2}a_0^2
 \end{eqnarray}
 and 
 \begin{equation}
 [l_n,a_m] = -m a_{n+m}
 \end{equation}
Vertex operators $V_{\beta} = :e^{i\beta \varphi(z)}:$ are $\textrm{H}$ primaries:
\begin{eqnarray}
& & a_n V_{\beta}= 0  \qquad n > 0 \\
& & a_0 V_{\beta} = \beta V_{\beta}
 \end{eqnarray}
and as a consequence are also Virasoro primaries (for $l_n$) with conformal dimension $\frac{\beta^2}{2}$.

\subsubsection{Ward identities}

We are concerned with correlation functions:
\begin{equation}
F_{a,b} = \langle a | V(z_1) V(z_2) \cdots V(z_N) | b \rangle
\end{equation}
where $\langle \beta |$ and $| \alpha \rangle$ are fields (arbitrary descendants) at $\infty$ and $0$ respectively. We denote by $\overleftarrow{X_n}$ and $\overrightarrow{X_n}$ the action of the mode $X_n$ on these vectors: 
\begin{eqnarray}
\overleftarrow{X_{-n}} & = & \frac{1}{2\pi i} \oint_{\infty} dz z^{n+\Delta-1} X(z) \\
\overrightarrow{X_n} & = & \frac{1}{2\pi i} \oint_{0} dz z^{n+\Delta-1} X(z)
\end{eqnarray}
where $\Delta$ is the (integer) dimension of the current $X(z)$. When acting on $\langle a | V(z_1) V(z_2) \cdots V(z_N) | b \rangle$, contour deformation yields the generic relation:
\begin{equation}
\overleftarrow{X_{-n}} = \sum_i \sum_{m=1-\Delta}^0 \binom{n+\Delta-1}{m+\Delta-1} z_i^{n-m} X_{m}^{(i)} + \overrightarrow{X_n}
\label{act_modes}
\end{equation}
where $X_m^{(i)}$ means that the mode $X_m$ acts on the field $V(z_i)$ in the correlator.
This leads to 
\begin{equation}
\overleftarrow{T_{-n}} =  \sum_{i} \left[ (g-1/2)(n+1)z_i^n + z_i^{n+1}\partial_i \right]+  \overrightarrow{T_n} 
\label{tnisomo}
\end{equation}
including the modified Ward identities:
\begin{eqnarray}
 \overleftarrow{T_{-1}} & = & \sum_i \left[ (2g-1) z_i + z_i^2 \partial_i  \right] + \overrightarrow{T_1} \\
  \overleftarrow{T_{0}} & = & \sum_i \left[ (g-1/2) + z_i \partial_i  \right] + \overrightarrow{T_0} \\
  \overleftarrow{T_{1}} & = & \sum_i  \partial_i  + \overrightarrow{T_{-1} }
\end{eqnarray}
and for the current $J$:
\begin{equation}
\overleftarrow{a_{-n}} = \sqrt{g/2} \sum_i z_i^n + \overrightarrow{a_n} 
\label{anisomo}
\end{equation}

\subsubsection{Contour deformations}

Using standard contour deformation techniques, we obtain for $n \in \mathbb{Z}$
\begin{equation}
a_{-n}^{(i)} =  -  \sum_{j (\neq i)}  \frac{\sqrt{g/2} }{(z_j-z_i)^n} -   (-1)^n \sum_{m \geq 0 } \binom{m+n-1}{n-1} z_i^{-m-n} \overrightarrow{a_m} +    \sum_{m \geq n } \binom{m-1}{n-1} z_i^{m-n}\overleftarrow{a_{m}}  \label{an_action_generic2}
\end{equation}
in particular
\begin{align}
a_{-1}^{(i)} & =  \sum_{j (\neq i)}    \frac{\sqrt{g/2}}{(z_i-z_j)}  +   \sum_{m \geq 0 } z_i^{-m-1}\overrightarrow{a_m} +  \sum_{m \geq 1 } z_i^{m-1} \overleftarrow{a_m}  \\
a_{-2}^{(i)} & = - \sum_{j (\neq i)}   \frac{\sqrt{g/2} }{(z_i-z_j)^2} - \sum_{m \geq 0 } (m+1)z_i^{-m-2} \overrightarrow{a_m} + \sum_{m \geq 2} (m-1)z_i^{m-2} \overleftarrow{a_m}\\
\end{align}
For the Virasoro mode $T_{-2}$:
\begin{equation}
T_{-2}^{(i)} =  \sum_{j (\neq i)} \left[  \frac{g-1/2}{(z_i-z_j)^2} + \frac{\partial_j}{(z_i-z_j)}  \right] +    \sum_{m \geq -1 } \frac{\overrightarrow{T_m}}{z_i^{m+2}} +\sum_{m \geq 2 } z_i^{m-2}\overleftarrow{T_m} 
\end{equation}

\subsection{Derivation of the correspondence}

\subsubsection{Correspondence at level 1}

The first order CS Hamiltonian is the generator of dilatations $H_1 = \sum_{i} (z_i \partial_i )$. Scale invariance dictates
\begin{equation}
\sum_{i} \left( \Delta + z_i \partial_i \right) F_{a,b} =    \left( \overleftarrow{T_0} - \overrightarrow{T_0}  \right) F_{a,b}
\end{equation}
and we get
\begin{equation}
I^{\pm}_2 =   T_0 = L_0 + \sum_{m>0} a_m a_{-m}  + \frac{1}{2} a_0^2 
\end{equation}
which is nothing but the zero mode of the total stress energy tensor in $Vir (g)\otimes \textrm{H}$. Since $a_0$ commutes with the whole algebra, we are free to choose 
\begin{equation}
I^{\pm}_2 =   L_0 + \sum_{m>0} a_m a_{-m}
\end{equation}

\subsubsection{Correspondence at level 2}

The correspondence between the order $2$ CS Hamiltonian 
\begin{equation}
H_2^g=\sum_{i=1}^N \left(z_i \frac{\partial}{\partial {z_i}}\right)^2 + g(1-g)\sum_{i\neq j} \frac{z_iz_j}{z_{ij}^2}
\end{equation} 
and the operator $I_3$ comes from the degeneracy at level $2$ in the module of $V = \Phi_{(1|2)} :  e^{ i \sqrt{\frac{g}{2}} \varphi}:$
\begin{equation}
\left(L_{-1}^2 - g L_{-2} \right)V=0
\end{equation}
An important point is that $L_{-1}$ is no longer a simple derivative in the extended algebra $Vir (g)\otimes \textrm{H}$, as we must now deal with the total Virasoro algebra generated by  $T_n = L_n+l_n$. This degeneracy becomes:
\begin{equation}
\left( T_{-1}^2  - g T_{-2}    + (g-1)\sqrt{g/2}a_{-2} +  g a_{-1}^2  - 2\sqrt{g/2}a_{-1}T_{-1}\right) V = 0
\end{equation}
This yields the following relation:
\begin{align}
\sum_{i=1}^N z_i^2\left( \partial_i^2  - g T^{(i)}_{-2}    + (g-1)\sqrt{g/2}a^{(i)}_{-2} +  g \left(a^{(i)}_{-1} \right)^2  - 2\sqrt{g/2}a^{(i)}_{-1}\partial_i \right) F_{a,b} = 0
\end{align}
Using the contour deformations of the previous section all these terms can be handled. For instance
\begin{align}
\sum_{i=1}^N z_i^2 T_{-2}^{(i)} \langle \beta | V(z_1) V(z_2) \cdots V(z_N) | \alpha \rangle =  \left[  \sum_{m \geq 0} p_{-m} \overrightarrow{T_m} + \sum_{ m \geq 1} p_m \overleftarrow{T_m} \right] \langle \beta | V(z_1) V(z_2) \cdots V(z_N) | \alpha \rangle \\ 
+ \left[  \left(g-\frac{1}{2}\right) \sum_{i \neq j} \frac{z_i^2}{z_{ij}^2} -\frac{1}{2} \sum_{i \neq j} \frac{z_iz_j}{z_{ij}} (\partial_i-\partial_j)  - \sum_{i=1}^N z_i\partial_i\right] \langle \beta | V(z_1) V(z_2) \cdots V(z_N) | \alpha \rangle 
\end{align}
We obtained the following relation for $F_{a,b} =   \langle a | V(z_1) V(z_2) \cdots V(z_N) | b \rangle $
\begin{align}
H_{2}^g F_{a,b}=  \left[ \overrightarrow{I_3^{(-)}}(g) + \overleftarrow{I_3^{(+)}}(g) + 4(g-1) \sum_{m\geq 1}m \overleftarrow{a_m}\overrightarrow{a_m}  + \mathcal{E}{}\right] F_{a,b} \label{generic_CS}
\end{align}
with
\begin{equation}
I_3^{(\pm)}(g) = 2(1-g)\sum_{m\geq 1} m a_{-m}a_m  \pm \sqrt{2g} \sum_{m\geq 1} \left( a_{-m} L_{m} + L_{-m}a_m \right) \pm  \sqrt{\frac{g}{2}}\left(  \sum_{m,k \geq 1} a_{-m-k}a_m a_k  +  a_{-m}a_{-k}a_{m+k} \right)
\label{I3virp}
\end{equation}
The extra term $\sum_{m\geq 1}m \overleftarrow{a_m}\overrightarrow{a_m} $ vanishes identically whenever $a$ or $b$ is primary, and $\mathcal{E}$ is simply a constant:
\begin{eqnarray}
\label{const_MM}
\mathcal{E} F_{a,b} &=& \left[ g(N-1) + (1-g) +2\sqrt{\frac{g}{2}}\overrightarrow{a_0} \right] \left( \sum_{i} z_i \partial_i \right)F_{a,b}  \\ \nonumber
& +& \left[   gN\overrightarrow{T_0}  + (g-1)N\sqrt{\frac{g}{2}}\overrightarrow{a_0} - g \left( g \binom{N}{3} + (1-g)\binom{N}{2} +2\sqrt{\frac{g}{2}}\overrightarrow{a_0} \binom{N}{2} + N\overrightarrow{a_0}^2 \right) \right ]F_{a,b} 
  \end{eqnarray}

\subsubsection{Correspondence at level 3}

It is quite natural to expect that a relation involving the order $3$ CS Hamiltonian
\begin{equation}
H_3^g=\sum_{i=1}^N \left(z_i \frac{\partial}{\partial {z_i}}\right)^3 + \frac{3}{2}g(1-g)\sum_{i\neq j} \frac{z_iz_j}{z_{ij}^2}( z_i \partial_i - z_j \partial_j)
\end{equation} 
can be obtained from a degeneracy at level $3$ of $V = \Phi_{(1|2)} :  e^{ i \sqrt{\frac{g}{2}} \varphi}: $. However there are two such null states: 
\begin{itemize}
\item $L_{-1} \left(L_{-1}^2 - g L_{-2} \right)V=0 $
\item $a_{-1}\left(L_{-1}^2 - g L_{-2} \right)V=0 $
\end{itemize}
and taking a generic linear combination of them will not work:  the corresponding relation will not separate into a differential operator on one side and an operator acting in the conformal Hilbert space on the other side. It turns our that demanding this separability amounts to consider the degeneracy:
\begin{equation}
\left( L_{-1} + 3\sqrt{\frac{g}{2}} a_{-1} \right) \left(L_{-1}^2 - g L_{-2} \right)V=0
\end{equation}
i.e.
\begin{equation}
T_{-1}^3 - g T_{-1}T_{-2} + (g-3)\sqrt{\frac{g}{2}}a_{-2}T_{-1} - g a_{-1}^2T_{-1} - g \sqrt{2g} a_{-1}T_{-2}+ (g-1)\sqrt{2g}a_{-3} +g(g+1)a_{-1}a_{-2} + g\sqrt{2g}a_{-1}^3
\end{equation}
Using the same techniques as for the level $2$ degeneracy we obtained the expression
\begin{align}
\label{Iquatre}
I^{\pm}_4(g) & = -g \sum_{m >0 } L_{-m}L_{m}    \\ \nonumber
& -\frac{3}{2}g \sum_{m,p >0 } \left(  2L_{-p}a_{-m}a_{p+m} + 2 a_{-m-p}a_{m}L_{p} + a_{-m}a_{-p}L_{p+m} + L_{-m-p}a_{m}a_{p} \right)   \\ \nonumber
& \pm \frac{3}{2}\sqrt{2g}(g-1) \sum_{m >0 } m(a_{-m}L_m + L_{-m}a_{m}) \pm 3\sqrt{2g}(g-1) \sum_{m,p >0 } m ( a_{-m}a_{-p}a_{m+p} + a_{-m-p}a_ma_p)  \\ \nonumber
& -\frac{1}{2}g L_0^2 - 3g L_0 \sum_{m>0} a_{-m}a_m  +\sum_{m\geq 1}\left[ \frac{1}{2} ( 9g-5-5g^2)m^2 - \frac{1}{2} (g-1)^2   \right] a_{-m}a_m
\\ \nonumber
  &- \frac{g}{8} \sum_{\substack{m_1+m_2+m_3+m_4 = 0 \\ m_i \neq 0}} :a_{m_1}a_{m_2}a_{m_3}a_{m_4}: \label{I4}
\end{align}
\newsec{Integral-differential representation of non-polynomial Calogero Sutherland eigenfunctions }

We have shown that   the  correlation functions (\ref{qhMM}) of CFT minimal models  admit a particular  separation of variables, see (\ref{dual_exp_MM}). This result   points out  the existence of a  new  families of eigenfunctions of (\ref{CS_g_g}), the $\mathcal{F}^{\alpha,a}_{\lambda}(z)$. 

Each of these solutions is associated to a  set   $\lambda$ which is related to  the two  partitions  $n^{e}$
and  $n^o$,  $\lambda \to \left[n^{e}, n^{o}\right]$  (\ref{two_young}).  The associated  eigenvalue have been 
given in  (\ref{dual_energy-MM}). 

Here we  want to show that:
\begin{itemize}

\item the function $\mathcal{F}^{\alpha,a}_{\lambda^0}(z)$,  see   (\ref{dual_exp_MM}) and  (\ref{def_lambda_max_g}),   can be expressed in terms of contour integrals by using standard Coulomb gas methods.

\item  the "excited states"  $\mathcal{F}^{\alpha,a}_{\lambda}(z)$ can be written in terms of symmetric differential operators acting on $\mathcal{F}^{\alpha,a}_{\lambda^0}(z)$.

\end{itemize}

\underline{Coulomb gas representation of $\mathcal{F}^{\alpha,a}_{\lambda^0}(z)$}

Consider for instance  the  eigenfunction  $\mathcal{F}^{\alpha,b}_{\lambda^0}(z)$, $\alpha^{-1}=1-g$,       constructed from the  conformal block of   $\Phi_{(1|2)}$ primary fields:  
\begin{equation}
\mathcal{F}^{1/(1-g),b}_{\lambda^0}(z)=\langle \Phi_{(1|2)}(z_1)\dots \Phi_{(1|2)}(z_N)\rangle_{b} \prod_{1\leq i<j\leq N} z_{ij}^{2h}
\label{min_ener}
\end{equation}
 where  the dimension $h$  is given in  (\ref{conf_dim_g}).
 Note that, with respect to the definition (\ref{fap}) given in section (\ref{corr}),  the function $\mathcal{F}^{1/(1-g),b}_{\lambda^0}(z)$ corresponds to :
 \begin{equation}
 \mathcal{F}^{1/(1-g),b}_{\lambda^0}(z)=\langle e^{i N \sqrt{g/2} \varphi}\otimes \mbox{I} | V(z_1) V(z_2) \cdots V(z_N) | \mbox{I}\otimes \mbox{I}\rangle_b \prod_{1\leq i<j\leq N} z_{ij}^{g-1}.
 \end{equation} 
 In the above equation, the factor $\prod_{1\leq i<j}  z_{ij}^{g-1} 
$ can also be understood by remembering  the fact that  in section (\ref{corr}) we worked  in the gauge of the Hamiltonian (\ref{cs_gamma}) with $\gamma=0$ while here we are considering the eigenfunctions of (\ref{cs_gamma}) with $\gamma=1-g$. 
  
In the Coulomb gas representation, the operators $\Phi_{(1|2)}(z_1)$ are represented by  vertex operator, see (\ref{vert_op_b}). A crucial role is played then by the screening operators:
\begin{equation}
V_{\pm}(z)=e^{i\alpha_{\pm} \phi(z)}
\end{equation}
which are primary operator of conformal dimension $h_{\pm}=1$.  
Suppose one in interested in the computation of a general conformal block 
\begin{equation}
\langle \prod_i^N \Phi_{\alpha_i}(z_i)\rangle.
\end{equation}
The above conformal block is obtained by considering  the following free boson correlation function
\begin{eqnarray}
\Phi(z,x,y)&\equiv& \langle \prod_i e^{i \alpha_i \phi(z_i)}\prod_j^n V_{+}(x_j)\prod_j^m V_{-}(y_j)\rangle=\\
&&=\prod_{1\leq i<j\leq N} z_{ij}^{2\alpha_i \alpha_j} \prod_{1\leq i \leq N}\prod_{1\leq j\leq n} (z_i-x_j)^{2\alpha_{+}\alpha_{i}} \prod_{1\leq i \leq N}\prod_{1\leq j\leq m} (z_i-y_j)^{2\alpha_{-}\alpha_{i}} \\ &&
 \prod_{1\leq i<j\leq n} (x_i-x_j)^{2\alpha_{+}^2}\prod_{1\leq i \leq n}\prod_{1\leq j\leq m} (x_i-y_j)^{-2},
\end{eqnarray}
where the number of screenings $n$ and $m$ is chosen in order to satisfy the charge neutrality condition:
\begin{equation}
 \sum_i \alpha_i+n \alpha_{+}+m \alpha_{-}=2\alpha_0,
\label{ch_neu}
\end{equation}
  with $\alpha_0$  given in (\ref{min_mod_param}).  The conformal block  $\langle \prod_i^N \Phi_{\alpha_i}(z_i)\rangle$  is given by integrating the   $\Phi(z,x,y)$ over the positions of the $n+m$ screenings. The  contours of integration have to be  closed in the Riemann surface on which  $\Phi(z,x,y)$ is defined. One has:
  \begin{equation}
  \langle \prod_i^N \Phi_{\alpha_i}(z_i)\rangle \propto \prod_i^n  \prod_{j}^m  \oint_{C_i} d x \oint_{S_j}d y \;\; \Phi(z,x,y)
  \end{equation}

   By setting $\Phi_{12}(z_i)= e^{i \alpha_{12} \phi(z_i)}$, for $i=1,2,\dots ,N-1$ and   $\Phi_{12}(z_N)= e^{i (2\alpha_0-\alpha_{12}) \phi(z_N)}$,  the charge neutrality condition (\ref{ch_neu}) associated to the conformal block in (\ref{min_ener}) is satisfied with $n=(N-2)/2$. We have then:
   \begin{equation}
\mathcal{F}^{1/(1-g),b}_{\lambda^0}(z)   \propto  \prod_{1\leq i<j\leq N} z_{ij}^{\alpha_{+}^2/2} \prod_{1\leq i\leq (N-2)/2}    \oint_{C_i} dx  \prod_{1\leq i\leq N}\prod_{1\leq j \leq (N-2)/2}  (z_i-x_j)^{-\alpha_{+}^2} \prod_{1\leq i<j\leq (N-2)/2}  (x_i-x_j)^{2\alpha_{+}^2},
  \end{equation}
 Note that one could directly  prove that a function of the above form defines, for appropriate closed integration contours $C_i$,  a solution of (\ref{CS_g_g}) (see for instance \cite{Kaneko,Dubedat})  

The different conformal blocks giving  $\mathcal{F}^{1/(1-g),b}_{\lambda^0}(z)$ (we recall that in  our notation the different conformal blocks  are  indexed by the integer $b$) correspond to  the different independent ways of  choosing closed contours $C_i$.  One can easily show (\cite{Cardy_Doyon}) that each choice is associated to certain boundary conditions. As an exemple, we can let the contour $C_i$ encircling the points $z_{2 i-1}$ and $z_{2 i}$ (with figure-8 contours in order to take into account the branch cuts).  The contours can then be shrinked  to give the following integral expression:
 \begin{equation}
\mathcal{F}^{1/(1-g),1}_{\lambda^0}(z)   \propto  \prod_{1\leq i<j\leq N} z_{ij}^{\alpha_{+}^2/2} \prod_{1\leq i\leq (N-2)/2}    \int_{z_{2 i-1}}^{z_{2 i}} d x_i  \prod_{1\leq i\leq N}\prod_{1\leq j \leq (N-2)/2}  (z_i-x_j)^{-\alpha_{+}^2} \prod_{1\leq i<j\leq (N-2)/2}  (x_i-x_j)^{2\alpha_{+}^2}.
  \end{equation}
The $\mathcal{F}^{1/(1-g),1}_{\lambda^0}(z)$ corresponds to the conformal block where the fields $\Phi(z_{2i-1})$ and $\Phi(z_{2i})$ fuse into the identity channel: the function  $\mathcal{F}^{1/(1-g),1}_{\lambda^0}(z)$ behaves has  $\mathcal{F}^{1/(1-g),1}_{\lambda^0}(z)\sim (z_{2i-1}-z_{2i})^{0}$ for $z_{2i} \to z_{2i -1}$.

For general central charge,  the number of conformal block is the  Catalan of $N/2$,  $C_{N/2}$,i.e. $b=1,\dots ,C_{N/2}$. Naturally for values of central charge associated to rational CFTs,   one has  to take into account the truncations in the operator algebra. This is the case for instance for the $c=1/2$ theory, which we have discussed in detail in section (\ref{Ising_cs}).   

It is interesting to mention  that   in  \cite{ItOo} it was shown how to derive   the (anharmonic ratio-) expansion of a four-point conformal block  from its  Coulomb gas representation. 
 
   \underline{Formal solutions for  the excited states.}
   
   It is interesting to  recall a known result  for the Jack eigenfunctions $J^{\alpha}_{\lambda}(z)$.  One defines  \cite{LPV}  a class of operators $\mathcal{D}_{\lambda}$  which are particular symmetric combination of differential  operators acting on a  function $f(z)$,  $\mathcal{D}_{\lambda}[f(z)]$,   and  which are associated to a partition $\lambda$. The Jack polynomials can be obtained acting with this class of symmetric operators on the ground state wavefunction  $J^{\alpha}_{\lambda^0}(z)$,   with $\lambda^0=[\emptyset]$, which is the trivial constant function,  $J^{\alpha}_{[\emptyset]}(z)=1$:
\begin{equation}
J^{\alpha}_{\lambda}(z)=\mathcal{D}^{\alpha}_{\lambda}[J^{\alpha}_{[\emptyset]}(z)]=\mathcal{D}^{\alpha}_{\lambda}[1]
\label{oper_solut}
\end{equation}

 In an analogous way,  we can show that the non-polynomial solution $\mathcal{F}^{1/(1-g),b}_{\lambda}(z)$ can be obtained  by acting with a symmetric differential operator $\mathcal{O}_{\lambda}[f(z)]$ on the reference state $\mathcal{F}^{1/(1-g),b}_{\lambda^0}(z)$: 
\begin{equation}
\mathcal{F}^{1/(1-g),b}_{\lambda}(z)=\mathcal{O}_{\lambda}[ \mathcal{F}^{1/(1-g),b}_{\lambda^0}(z)]
\label{gen_res}
\end{equation}

We have seen in section (\ref{corr}) that an eigenfunctions of $H_2^g=H^{g,0}$, see (\ref{cs_gamma}),  can take the form (\ref{def_F}). We consider here the function $F^{(g,+)}_{\lambda}(z)$, see (\ref{def_F}), in which the  primary  $|P>$ and the state  $|P^{+}_{\lambda}(g)\rangle$ appearing in  (\ref{def_F}) are respectively the trivial state $\mbox{I}\otimes \mbox{I}$ and a descendant state of   $ e^{i N \sqrt{g/2} \varphi}\otimes \mbox{I}$:
\begin{equation}
F^{(g,+)}_{\lambda}(z)=\langle  P^{+}_{\lambda}(g)| V(z_1) V(z_2) \cdots V(z_N) | \mbox{I}\otimes \mbox{I}\rangle_b.
\end{equation}
We recall  that $|P^{+}_{\lambda}(g)>$  is  an eigenstate of $I_{n+1}^{(+)}(g)$ and  it is labelled  by  the two partitions $[n^{e}, n^{o}]$, $P^{+}_{\lambda}(g) \to P^{+}_{([n_e],[n_o])}(g)$, see sections (\ref{minmodg1})-(\ref{minmodggen}).  
Again, if we are interested in the  eigenfunctions of the CS Hamiltonian $H^{g,1-g}$ we have simply:
\begin{equation}
\mathcal{F}^{1/(1-g),b}_{\lambda}(z)=  F^{(g,+)}_{\lambda}(z) \prod_{1\leq i<j\leq N} z_{ij}^{g-1}.
\end{equation}

 In order to simplify the notation, we indicate  $|\mbox{I}\otimes \mbox{I}>\equiv|0>$, $|e^{i N \sqrt{g/2} \varphi}\otimes \mbox{I}>\equiv|N>$ and $|P^{+}_{([n_e], [n_o])}(g)\rangle\equiv|[n_e],[n_o]>$

One can then show the validity of (\ref{gen_res}) by the following procedure:
\begin{itemize}

\item  Find the basis   $|[n_e],[n_o]\rangle$ which diagonalizes the operator $I_{3}^{+}(g)$(\ref{I3}).  Each  state  $|[n_e],[n_o]\rangle$ is  then found  as a  given  combination $\mathcal{O}_{\lambda}(\{L_{-n}\},\{a_{-n}\})$ of Virasoro modes $L_{-n}$  and of $u(1)$ current modes $a_n$ acting on $ e^{i N \sqrt{g/2} \varphi}\otimes \mbox{I}$ state:

\item From the relations ($\ref{act_modes}$), one can  write the action of the algebra modes in terms of differential operator acting on the  correlation function $F^{(g,+)}_{([\emptyset],[\emptyset])}(z)=<N| V(z_1) V(z_2) \cdots V(z_N) | 0\rangle_b$:
\begin{equation}
F^{(g,+)}_{([n_e],[n_o])}(z)=\mathcal{O'}_{([n_e],[n_o])}[F^{(g,+)}_{([\emptyset],[\emptyset])}(z)]
\end{equation}
where $\mathcal{O'}_{([n_e],[n_o])}[f(z)]$ will be particular symmetric combination of differential operators, associated to the descendant $|a_{\lambda}>$, acting on a function $f(z)$.

\item The differential operator defined in (\ref{gen_res}) is then
\begin{equation}
 \mathcal{O}_{([n_e],[n_o])}= \left(\prod_{1\leq i<j\leq N} z_{ij}^{1-g}\right)\mathcal{O'}_{([n_e],[n_o])}\left(\prod_{1\leq i<j\leq N} z_{ij}^{g-1}\right)
\end{equation}
\end{itemize}

We  illustrate the above  procedure  by considering  the basis $|[n_e],[n_o]>$  which diagonalizes $I_3^{g}$, see (\ref{I3vir}),  at the first and at the second level in the module  $|N>$. 

\underline{Descendants level one}

At the first level there is only one descendant. 
\begin{equation}
|[1],[\emptyset]>=\mathcal{O'}_{|[1],[\emptyset]}(\{L_{-n}\},\{a_{-n}\})  |0>=a_{-1}|N>
\end{equation} 
Using (\ref{anisomo}) one has:
\begin{equation}
\langle  [1],[\emptyset]| V(z_1) V(z_2) \cdots V(z_N) | 0\rangle_b=
\left(\sum_{i} z_i\right) \langle  N| V(z_1) V(z_2) \cdots V(z_N) |0>.
\end{equation}
 This means that the operator $\mathcal{O'}_{|[1],[\emptyset]}[f(z)]$ simply  multiplies the function $f(z)$
 by the monomial $m_{[1]}(z)$: 
\begin{equation}
\mathcal{O'}_{|[1],[\emptyset]}[f(z)]=\sum_{i} z_i f(z)=m_{[1]}(z)f(z).
\end{equation} 

\underline{Descendants level two}

We have three descendants at level two.  In the basis  $a_{-2}|0>$ and $\sqrt{2g} a_{-1}^2|0>$ and $\sqrt{2g}L_{-2}|0>$, the operator  $I^{+}_{3}(g)$, reads
 
 \begin{center}
 $I^{+}_{3}(g)=\left(
 \begin{array}{ccc}
8(1-g) & 2 g & g \;c(g) \\
 1 & 4(1-g) & 0 \\
 2& 0 & 0
\end{array}\right)
$
\end{center}
The correspondent eigenvectors are:
\begin{eqnarray}
|[2],[\emptyset]>&=&\left[(3-2 g)a_{-2} +\left(\frac{3}{2}-g\right)\sqrt{2g} a_{-1}^2+\sqrt{2 g} L_{-2}\right] |N> \nonumber \\
|[1,1],[\emptyset]>&=&\left[(2-3 g)a_{-2} +\left(\frac{3}{2}-\frac{1}{g}\right)\sqrt{2g} a_{-1}^2+\sqrt{2 g} L_{-2}\right] |N>\\
|[1],[1]>&=&\left[(1-g)a_{-2} -\frac{1}{2} \sqrt{2g} a_{-1}^2+\sqrt{2 g} L_{-2}\right] |N> \nonumber
\end{eqnarray}
with eigenvalues  $6-4 g$, $4-6 g$ and $2 -2g$ respectively. Note that for $g=1$ this is  the same basis as in Section \ref{minmodg1} where we used instead the bosonic representation of the Virasoro algebra. 
Taking into account:
\begin{equation}
 L_{-2}|N>=(T_{-2}-a_{-2}a_{0} -\frac{1}{2}a_{-1}^2)|N>
\end{equation}
it is  straightforward to use the relations (\ref{tnisomo}) and (\ref{anisomo}) to associate to each descendant a symmetric differential operator acting on the function $F^{(g,+)}_{([\emptyset],[\emptyset])}(z)$. By remplacing:
\begin{eqnarray}
L_{-2}|N>&\to& \left[ \left(\frac{11-2 N}{4} g -\frac{1}{2}\right)m_{[2]}(z) -\frac{g}{2} m_{[1,1]}+\sum_{1\leq i\leq N} z_i^3\partial_i \right]F^{(g,+)}_{([\emptyset],[\emptyset])}(z) \\
a_{-2}|N>& \to&\left[\frac{g}{2} m_{[2]}(z)\right]F^{(g,+)}_{([\emptyset],[\emptyset])}(z)  \\
a_{-1}^2|N>& \to&\left[\frac{g}{2} m_{[2]}(z)+g m_{[1,1]}(z)\right]F^{(g,+)}_{([\emptyset],[\emptyset])}(z)
\end{eqnarray}
the  expressions for $\mathcal{O}^{'}_{[n_e],[n_o]}[f[z]]$ and for  $\mathcal{O}_{[n_e],[n_o]}[f[z]]$ are easily found. In appendix B, where we considered the case with $N=4$ and $g=4/3$, the    explicit expression for $\mathcal{O}_{[1],[1]}$ and $\mathcal{O}_{[1,1],[\emptyset]}$ were shown.


\begin{thebibliography}{10}

\bibitem{Ca}
F.~Calogero,
\newblock  J. Math. Phys. {\bf 10}, 2191, (1969).

\bibitem{Su}
 B.~Sutherland,
 \newblock  J. Math. Phys. {\bf 12}, 246 (1971); {\bf 12} , 251 (1971).
 
\bibitem{Ols_Per}
M.~A.~Olshanetsky and A.~M.~Perelomov,
\newblock  Phys. Rep. {\bf 94}, 313, (1983).

\bibitem{UHW}
H.~Ujino, K.~Hikami, M.~Wadati,
\newblock J. Phys. Soc. Japan, {\bf 61},  3425, (1992).

\bibitem{Poly}
A.~P.~Polychronakos,
\newblock Phys.
Rev. Lett. {\bf 69}, 703, (1992).

\bibitem{Dunkl_rev}
C.~F.~Dunkl,
\newblock  Trans. Am. Math. Soc. {\bf 311}, 167, (1989).

\bibitem{Dunkl_Opd}
E.~M.~Opdam,
\newblock arXiv:math/9812007.



\bibitem{Ha} Z. N. C.~Ha,
\newblock  Phys.~Rev.~Lett. {\bf 73}, 1574, (1994);
Nucl. Phys. B {\bf 435}, 604, (1995).


\bibitem{SLP}
D.~Serban, F.~Lesage, V.~Pasquier, 
\newblock Nucl. Phys. B {\bf 435}, 585, (1995); Nucl.~Phys.~B {\bf 466}, 499, (1996).

\bibitem{Forr}
P.~J.~Forrester,
\newblock Nucl. Phys. {\bf B 388}, 671, (1992).


\bibitem{AMOS}
H. Awata, Y. Matsuo, S. Odake, J. Shiraishi,
\newblock Phys. Lett. {\bf B 347}, 49, (1995).

 \bibitem{Kaneko}
J.~Kaneko,
\newblock
Siam J. Math. Anal. {\bf  24}, 1086, (1993). 

\bibitem{Stanley}
R.~Stanley,
\newblock Adv. Math. 77 (1989) 76-115.

\bibitem{Gaudin}
M.~Gaudin, 
\newblock Saclay preprint SPhT92-158.




\bibitem{Macdonald}
I.~G. Macdonald, 
\newblock {\it Symmetric functions and Hall polynomials}, 2nd ed., Oxford University
Press, New York, 1995.

\bibitem{MY}
K.~Mimachi and Y.~Yamada,
\newblock Commun. Math. Phys. {\bf 174}, 447, (1995).

\bibitem{SSDFR}
R.~Sakamoto, J.~ Shiraishi, D.~ Arnaudon, L.~ Frappat and  E.~ Ragoucy,
\newblock Nucl. Phys. {\bf B704}, 490, (2005). 

\bibitem{AMOS_w}
H.~Awata, Y.~Matsuo, S.~Odake, J.~Shiraishi,
\newblock Nucl. Phys. {\bf B449}, 347, (1995).

\bibitem{Cardy}
J.~Cardy,
\newblock Phys. Lett. {\bf B 582}, 121, (2004).

\bibitem{Cardy_Doyon}
B.~Doyon and J.~Cardy,
\newblock  J.~Phys.~A~{\bf 40}, 2509, (2007).


\bibitem{ES}
B. Estienne and R. Santachiara,
\newblock  J. Phys. A: Math. Theor. {\bf 42}, 445209, (2009). 

\bibitem{EBS}
B. Estienne, A.~Bernevig and  R. Santachiara,
\newblock  Phys. Rev. {\bf B 82}, 205307, (2010).

\bibitem{KJS}
K.~Schoutens
\newblock hys.Rev.Lett.{\bf 79}, 2608 (1997).

\bibitem{KJS_2}
E.~Ardonne, P.~Bouwknegt and K.~Schoutens
\newblock J.Stat.Phys. {\bf 102}, 421 (2001).

\bibitem{MooreRead}
G. Moore and N. Read,
\newblock Nucl. Phys. {\bf B 360}, 362 (1991); Prog. Theor. Phys. Suppl. {\bf 107} 157, (1992), hep-th/9202001. 

\bibitem{Nayak_Wilczek} 
C. Nayak and F. Wilczek,
\newblock Nucl. Phys. {\bf B 479}, 529 (1996).

\bibitem{ReadRezayi}
N. Read and E. Rezayi,
\newblock  Phys. Rev. {\bf B 59}, 8084 (1999).

\bibitem{AFLT}
V. A. Alba, V. A. Fateev, A. V. Litvinov and G. M. Tarnopolsky,  
\newblock Lett. Math. Phys. {\bf 98}, 33, (2011), arXiv:1012.1312 [hep-th].

\bibitem{AGT}
 L.~F.~Alday, D.~ Gaiotto and  Y.~Tachikawa, 
 \newblock  Lett. Math. Phys. {\bf 91},167, (2010). 

\bibitem{FL}
V.~A.~ Fateev and A.~V.~ Litvinov, 
 \newblock  JHEP {\bf 1002}, 014, (2010).

\bibitem{ST}
R.~Santachiara and A.~Tanzini,
 \newblock Phys. Rev. {\bf D 82}, 126006, (2010).

\bibitem{Nekrasov}
N. A. Nekrasov, 
\newblock Adv. Theor.
Math. Phys. {\bf 7}, 831, (2004), arXiv:hep-th/0206161.

\bibitem{ABW}
A. G. Abanov, E. Bettelheim, P. Wiegmann,
\newblock J. Phys.{\bf A 42}, 135201, (2009). 

 \bibitem{W}
N.~ Wyllard,
\newblock  JHEP {\bf 11}, 002,  (2009).

\bibitem{MM} 
A.~ Mironov and A.~ Morozov,
\newblock   Nucl. Phys. {\bf B825}, 1 (2010).

\bibitem{BB}
A. Belavin and V. Belavin,
\newblock Nucl. Phys. {\bf B 850}, 199, (2011).
 
\bibitem{Jevicki}
A.~Jevicki,
\newblock Nucl. Phys. {\bf B 376},  75, (1992).


 \bibitem{diFrancesco}
P. Di Francesco,  P. Mathieu and D. Senechal, 
\newblock {\it Conformal Field Theory}, Springer, NewYork (1997).


\bibitem{BGHP}
D ~Bernard, M.~Gaudin, F.D.M.~ Haldane and V.~ Pasquier,
\newblock J. Phys. {\bf A 26}  5219 (1993).

\bibitem{Haldane_Statistics}
F.~D.~M. Haldane,
\newblock Phys. Rev. Lett. {\bf67}, 937-940 (1991).

\bibitem{BernevigHaldane1}
B.~A. Bernevig and F.D.M. Haldane, 
\newblock Phys. Rev. Lett. {\bf 100}, 246802 (2008).

\bibitem{BernevigHaldane2}
B.~A. Bernevig and F.D.M. Haldane, 
\newblock Phys. Rev. Lett. {\bf 101}, 246806 (2008).

\bibitem{BernevigHaldane3}
B.~A. Bernevig and F.D.M. Haldane, 
\newblock Phys. Rev. Lett. {\bf 102}, 066802 (2009).

\bibitem{BernevigW}
B.~A. Bernevig, V. Gurarie, S.~H. Simon,
\newblock  J. Phys. A: Math. Theor. {\bf 42} 245206 (2009). 

 \bibitem{FJMM}
B. Feigin, M. Jimbo, T. Miwa and E. Mukhin,
\newblock International Mathematics Research Notices 1223 (2002).

\bibitem{FJMM2} 
B. Feigin, M. Jimbo, T. Miwa and E. Mukhin, 
\newblock International Mathematics Research Notices 1015 (2003); arXiv:math/0209042.

\bibitem{JM}
P.~Jacob and P.~Mathieu,
\newblock Phys. Lett. {\bf B627} 224 (2005).

\bibitem{ItOo}
H.~Itoyama and T.~ Oota,
\newblock 
Nucl.Phys.{\bf B838} 298 (2010). 

\bibitem{LPV}
L.~Lapointe and L.~Vinet,
\newblock Comm. Math. Phys. {\bf 178}, 425,  (1996). 

\bibitem{KMS} 
S.~Kanno, Y.~Matsuo and S.~Shiba,
\newblock Phys.Rev. {\bf D84}, 026007, (2011), arXiv:1105.1667.

\bibitem{BLZ}
V. Bazhanov, S. Lukyanov, A. Zamolodchikov,
\newblock 
Commun. Math. Phys. {\bf 177}, 381, (1996);
\newblock
Commun. Math. Phys. {\bf 190}, 247, (1997);
\newblock
Commun. Math. Phys. {\bf 200}, 297, (1999).

\bibitem{DOk}
D. Maulik and A. Okounkov, 
\newblock unpublished.

\bibitem{SWY}
B.~Shou, J.-F.~Wu, M.~Yu,
\newblock arXiv:1107.4784.

\bibitem{SHIRA}
H.~Awata, B.~Feigin, A.~Hoshino, M.~Kanai, J.~Shiraishi, S.~Yanagida,
\newblock Proceeding of RIMS Conference 2010 "Diversity of the Theory of Integrable Systems" (ed. Masahiro Kanai),
arxiv:1106.4088.

\bibitem{FL11}
V.~A.~Fateev, A.~L.~Litvinov,
\newblock arXiv:1109.4042.

\bibitem{Gurarie_is}
P.~Bonderson, V.~ Gurarie, C.~ Nayak,
\newblock Phys. Rev. {\bf B 83}: 075303, (2011).

\bibitem{Ardonne_is}
E.~Ardonne and  G.~Sierra
\newblock J. Phys. {\bf A 43}:505402, (2010).

\bibitem{DotsenkoFateev} 
Vl. S. Dotsenko and V.~A. Fateev,
\newblock Nucl. Phys. {\bf B324}, 312, (1984),
                         {\bf B251} 691(1985);
 \newblock Phys. Lett. {\bf B154}, 291, (1985).

\bibitem{Dubedat}
J.~Dubedat,
\newblock J. Stat. Phys. {\bf 123}, 1183 (2006).


\end{thebibliography}
\end{document}